\documentclass[intlimits,twoside,a4paper]{article}

\usepackage[cp1251]{inputenc}
\usepackage{bbding}
\usepackage[eqsecnum]{cmpj3}

\usepackage{bm}

\newcommand{\bea}{\begin{eqnarray}}
\newcommand{\eea}{\end{eqnarray}}
\newcommand{\be}{\begin{equation}}
\newcommand{\ee}{\end{equation}}
\newcommand{\eps}{\varepsilon}

\issue{2021}{24}{4}{43703}
\doinumber{10.5488/CMP.24.43703}

\title[Influence of external field direction on polarization rotation in antiferroelectric squaric acid]
{Influence of external field direction on polarization rotation in antiferroelectric squaric acid H$_2$C$_4$O$_4$ }

\author{A. P. Moina\orcid{0000-0002-2170-2445}}

\address{Institute for Condensed Matter Physics of the National Academy of Sciences of Ukraine,\\
	1 Svientsitskii St., 79011 Lviv, Ukraine}


\date{Received August 10, 2021, in final form October 2, 2021}
\begin{document}

\maketitle

\begin{abstract}
Using the previously developed model we explore the processes of polarization rotation in antiferroelectric crystals of squaric acid by the electric fields directed arbitrarily within the $ac$ plane.
Except for some particular directions of the field,
the two-step polarization reorientation at low temperatures
is predicted: first, to the noncollinear phase with perpendicular sublattice polarizations and then to the collinear ferroelectric phase. However, when the field is directed along the 
axis of spontaneous sublattice polarizations, the intermediate noncollinear phase is absent; when the field is at 45$^\circ$ to this axis, the field for transition to the ferroelectric phase tends to infinity. The ground state proton configurations and the directions of the sublattice polarization vectors are determined for all field orientations. The $T$--$E$ phase diagrams are constructed for the fields directed along the diagonals of the $ac$ plane and for the above discussed particular directions of the field.
\printkeywords

\end{abstract}


\maketitle


\section{Introduction}

The squaric acid H$_2$C$_4$O$_4$ is a classical two-dimensional antiferroelectric. The crystal is tetragonal,~$I4/m$, in the paraelectric phase and monoclinic,  $P2_1/m$, in the antiferroelectric one. The hydrogen bonded C$_4$O$_4$ groups form sheets, parallel to the $ac$ plane and stacked along the $b$-axis. Below the transition  at 373~K, a spontaneous polarization arises in these sheets, with the neighboring sheets polarized in the opposite directions \cite{semmingsen:95,semmingsen:77,hollander:77}. 
Protons on the hydrogen bonds in squaric acid move in double-well potentials, so each of the protons can occupy one of  the two sites on the bond: closer to the given C$_4$O$_4$ group or to the neighboring one. 
A double bond is formed between those two adjacent carbon atoms, closer to which two protons on the hydrogen bonds sit. The sublattice polarization is mostly electronic, while the direct contribution of ions and displaced protons is significantly smaller \cite{horiuchi:18}.

\begin{figure}[!h]
	\centerline{\includegraphics[width=0.7\columnwidth]{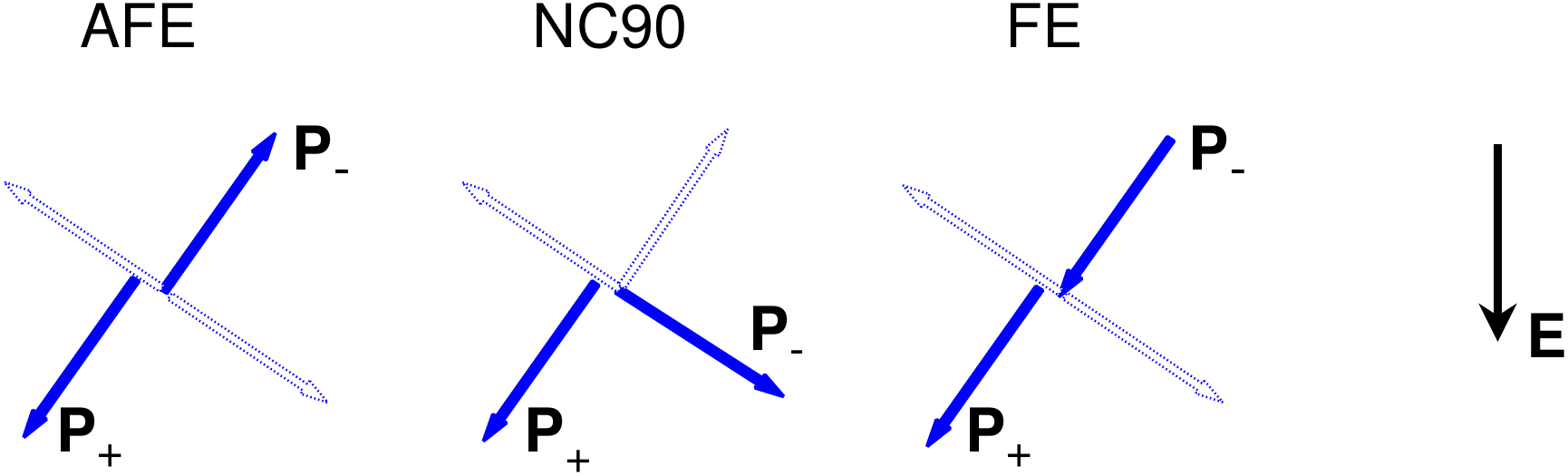}}
	\caption{(Colour online) The scheme of the two-step polarization rotation in squaric acid crystals at low temperature. Solid arrows show the actual  polarization vectors  $\mathbf{P}_{\pm}$ of the two sublattices in each phase: antiparallel in AFE, perpendicular in NC90, parallel in FE phase. Dotted arrows show the two other possible directions of these vectors, allowed by the crystal structure in the fully ordered system. The electric field $\mathbf{E}$ is noncollinear to the axes of sublattice polarizations. }
	\label{scheme}
\end{figure}

External electric field is known to switch a negative sublattice polarization in antiferroelectrics and induce thereby the transition from the antiferroelectric (AFE) to a ferroelectric (FE) phase, visualized in an experiment as the classical double $P-E$ hysteresis loops. 
In contrast to the uniaxial antiferroelectrics, where the
sublattice polarizations are collinear to one predetermined axis, the pseudo-tetragonal symmetry of the squaric acid crystal lattice and of its hydrogen bond networks allows the sublattice polarizations  in the fully ordered system to be directed along two perpendicular axes. As a result, in the AFE phase one of the sublattice polarizations can be rotated by 90$^\circ$ by the external field to go along the other allowed axis, at what a noncollinear ferrielectric phase with perpendicular sublattice polarizations (NC90 \cite{moina:21}) is induced. Hysteresis loop measurements and Berry phase calculations  \cite{horiuchi:18} have given evidence for such a rotation. Further calculations \cite{ishibashi:18} have indicated that the 90$^\circ$ rotation is possible at different orientations of the field within the $ac$ plane. It is also predicted \cite{horiuchi:18,ishibashi:18} that application of higher fields along the diagonals of the $ac$ plane can lead to the second rotation of the negative sublattice polarization by 90$^\circ$ and to the induction of the collinear ferroelectric phase. Experimentally, the transition to the FE phase in squaric acid has not been detected so far, because of the dielectric breakdown of crystal samples at the fields yet below 200~kV/cm \cite{horiuchi:18,Ol}. The general scheme of the sublattice polarization rotation and the sequence of the observed/predicted phases is given in figure~\ref{scheme}.

Recently, we developed the deformable two-sublattice proton-ordering model for squaric acid~\cite{moina:20} that described the effects associated with the diagonal lattice strains. Subsequently \cite{moina:21}, this model was  modified for description of the squaric acid behavior in external electric fields applied within the plane of the hydrogen bonds. The dipole moments associated directly with protons and with the $\pi$-bonds that can be switched by proton rearrangement were included into the model. The field-induced polarization rotation in this picture occurs via  flipping of one of the two protons in each molecule to the other site along the same hydrogen bond and a simultaneous switching of the $\pi$-bond to those carbon atoms, close to which the protons would sit after the flipping.

The temperature--electric field $T$--$E_1$ phase diagram of squaric acid has been constructed \cite{moina:21}  for the field $E_1$, directed along the  $a$ axis. The model calculations confirm the two-step process of polarization reorientation  \cite{horiuchi:18,ishibashi:18} at low temperatures, with the negative sublattice polarization being switched twice, by 90$^\circ$ at each transition. The system behavior is best characterized by the introduced noncollinearity angle $\theta$, which is the angle between the sublattice polarizations.  The collinear ferroelectric (FE) phase  and two noncollinear phases with almost antiparallel (AFE) and perpendicular (NC90) polarizations of the sublattices are identified. The diagram also contains the crossover region, where the noncollinearity angle varies continuously between, nominally, 180$^\circ$ and 90$^\circ$. The supercritical line, formed by the loci of the maxima of the dielectric permittivity isotherms, is used to mark the crossover between these two phases  \cite{moina:21}.

In the present paper, the model \cite{moina:21} is used without any further modifications to explore the polarization rotation in squaric acid by the electric fields of different directions within the $ac$ plane. In particular, it is interesting to check whether the field-induced phase transition to the collinear FE phase can be lowered down to more  experimentally realizable values of fields than it was predicted for the field $E_1$. Herein below, we give a brief outline of the model; all the details can be found in~\cite{moina:21}. In section~\ref{calculations} the low-temperature field direction--field magnitude phase diagram is constructed; the $T$--$E$  diagrams are presented for some specific field directions.

\section{The model}
\begin{figure}[!t]
	\centerline{\includegraphics[width=0.7\columnwidth]{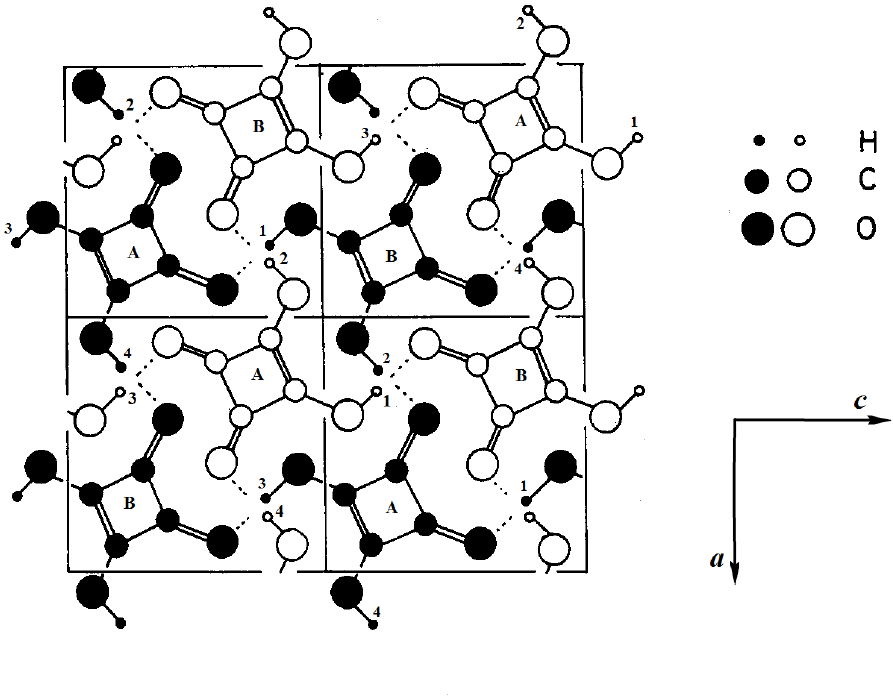}}
	\caption{Crystal structure of squaric acid as viewed along the $b$ axis. Figure is taken from \cite{semmingsen:74,moina:20,moina:21}. Two adjacent layers are shown, with black and open circles each. The A and B type C$_4$O$_4$ groups are indicated (see text for explanation), and the hydrogen bonds are numbered, $f=1,2,3,4$.} \label{sqa-structure}
\end{figure}

There are two formula units in the low-temperature monoclinic unit cell of squaric acid. In our model, the unit cell consists of two C$_4$O$_4$ groups and four hydrogen atoms ($f=1,2,3,4$, see figure~\ref{sqa-structure}) attached to one of them (the A type group).
All hydrogens around the B type groups are considered to belong to the A type groups, with which the B groups are hydrogen bonded.
Unless it is specified otherwise, the tetragonal axes $a$ and $c$
are used to identify the directions and angles within the  plane of hydrogen bonds.

The motion of protons in the double-well potentials is described by Ising pseudospins, whose two eigenvalues $\sigma_{yqf}=\pm 1$ are assigned to two equilibrium positions of the $f$-th proton. Here, $y$ stands for the layer index, and $q$ is the index of the A type C$_4$O$_4$ group. Two interpenetrating sublattices are formed by these layers with alternating polarization, to be denoted hereafter
as the plus and minus sublattices. It is assumed that the crystal is placed into electric fields $E_1$ and $E_3$ directed along the $a$ and $c$ axes.

These fields break the equivalence of the hydrogen bonds linking the C$_4$O$_4$ groups along the $a$ and $c$ axes (with $f=1,3$ and $f=2,4$).  Moreover, in presence of the fields the pseudospin mean values of the two sublattices differ not only by their signs, as in an antiferroelectric without the external field, but also by their absolute values. It means that we have four independent order parameters \cite{moina:21}
\[
\langle \sigma_{y q f} \rangle= \eta_{f\pm}\,,
\]
two for each sublattice
\begin{equation}
\label{symmetry1}
\eta_{1\pm}=-\eta_{3\pm}; \quad \eta_{2\pm}=-\eta_{4\pm}\,,
\end{equation}
the latter condition being imposed by the system translational symmetry. It should be mentioned that there is an experimental evidence for a slight non-equivalence of the hydrogen bonds, linking the C$_4$O$_4$ groups along the $a$ and $c$ axes, even in the absence of the field \cite{semmingsen:95}. This non-equivalence, however, is not taken into account by the present model, which is pseudotetragonal in the paraelectric phase.

The total system Hamiltonian \cite{moina:20,moina:21} includes ferroelectric intralayer long-range interactions, ensuring ferroelectric 
ordering within each separate layer, antiferroelectric interlayer interactions responsible for alternation of polarizations in the stacked layers, the short-range configurational interactions between protons, which also include  the interactions with external electric fields.

The short-range Hamiltonian describes the four-particle configurational correlations between protons sitting around each C$_4$O$_4$ group. The usual scheme \cite{matsushita:80,matsushita:82,moina:20,moina:21} of degenerate levels of four lateral (the energy ${\cal E}_a$), two diagonal (${\cal E}_s$), eight single-ionized with three or only one close protons (${\cal E}_1$), and two double-ionized  with four or zero protons  (${\cal E}_0$) configurations is assumed, with ${\cal E}_a<{\cal E}_s\ll{\cal E}_1\ll{\cal E}_0$. 
Only the groups with protons in the lateral and  single-ionized configurations have dipole moments in the $ac$ plane; the degeneracy of their energy levels is removed by the electric fields $E_1$ and $E_3$ (see  table~\ref{configurations_table}). 

\begin{figure}[!t]
	\centerline{
		\begin{tabular}{cc}
	\includegraphics[origin=c,angle=-90,width=0.18\textwidth]{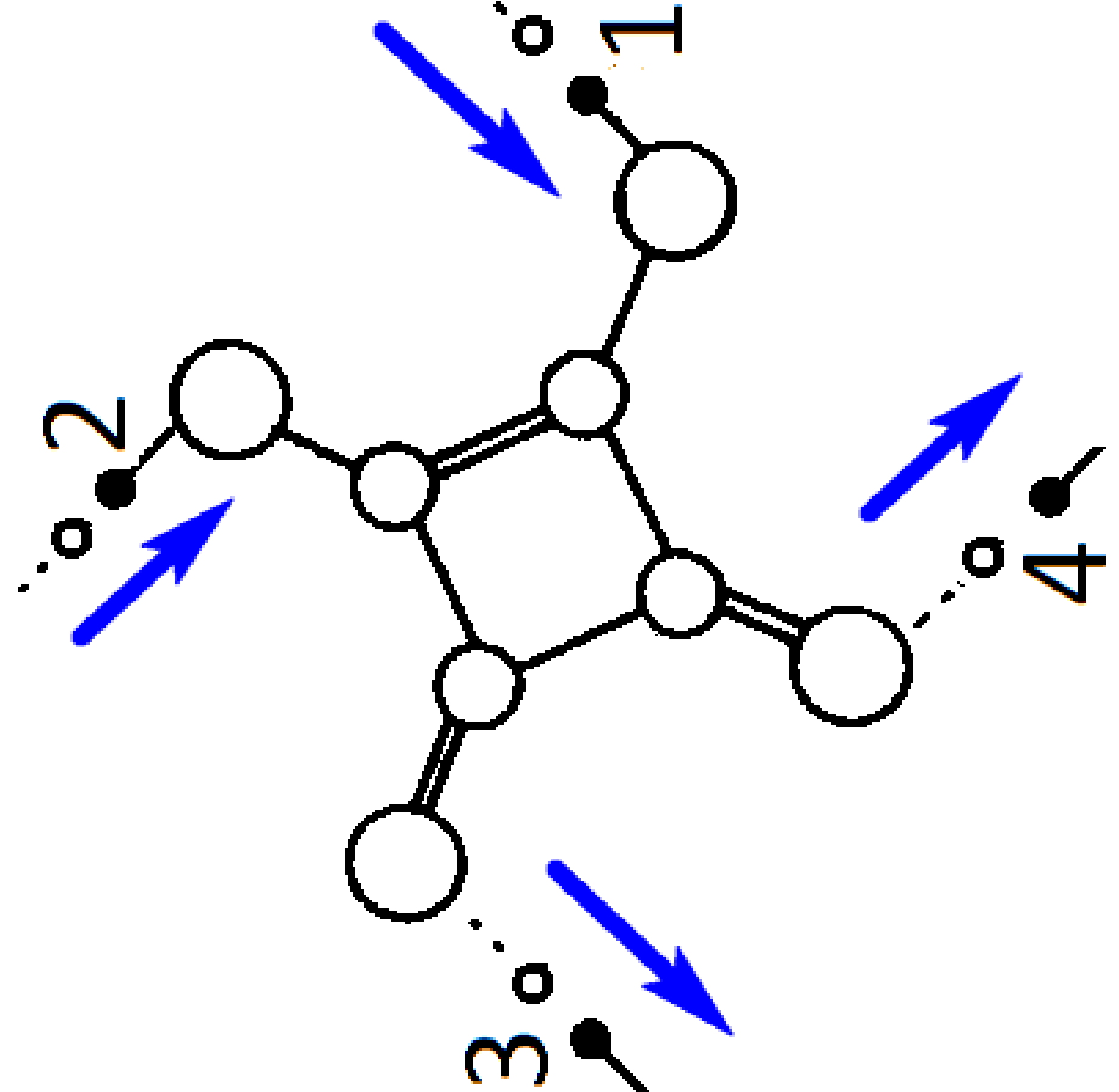}~~~
	\includegraphics[width=0.15\textwidth]{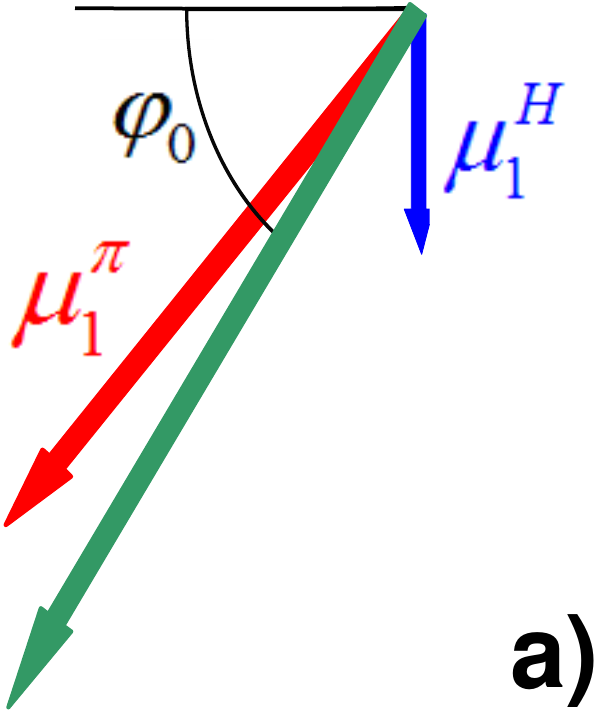}&
	\hspace{5ex}
\includegraphics[origin=c,angle=-90,width=0.18\textwidth]{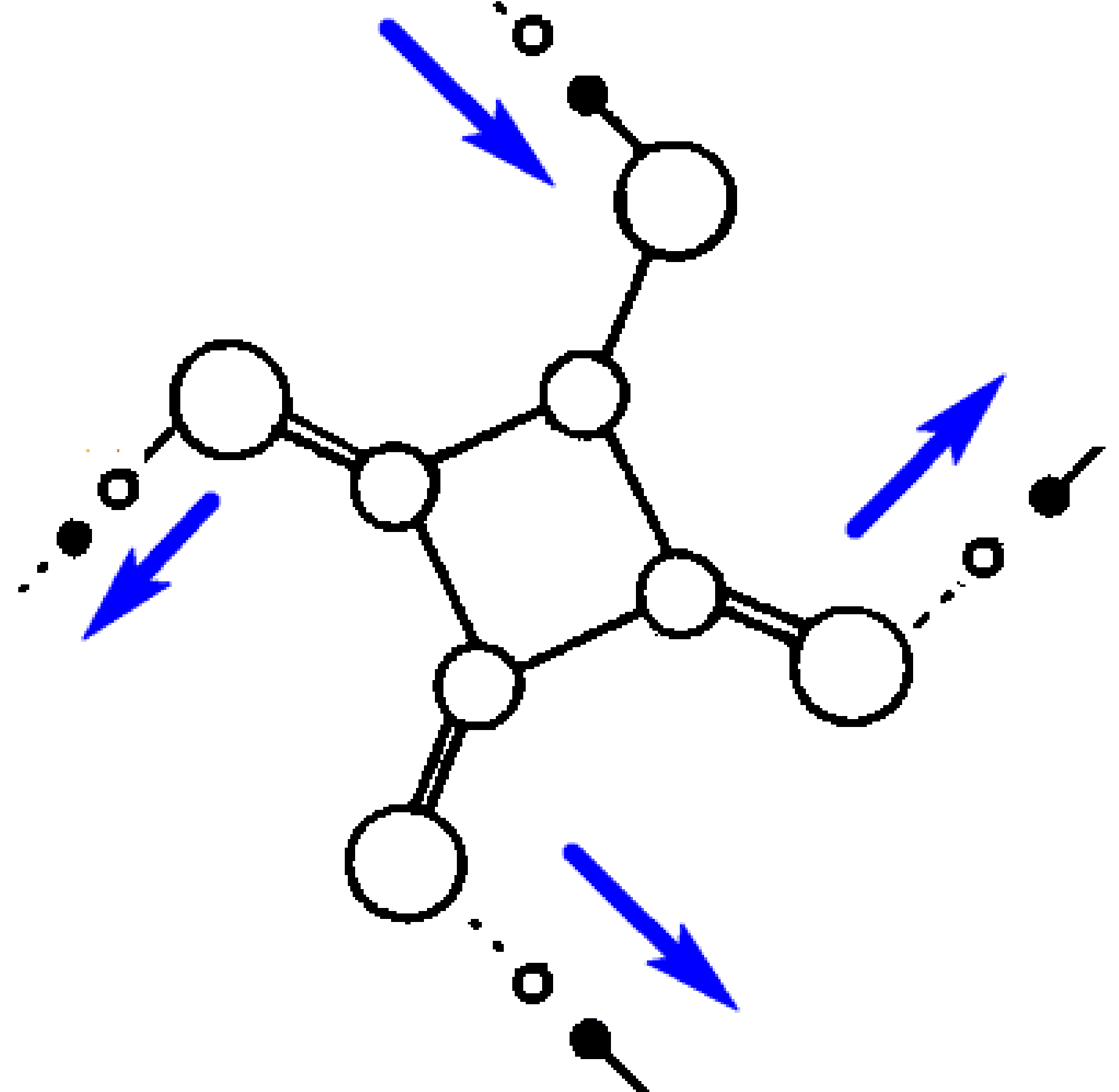}~~
	\includegraphics[width=0.22\textwidth]{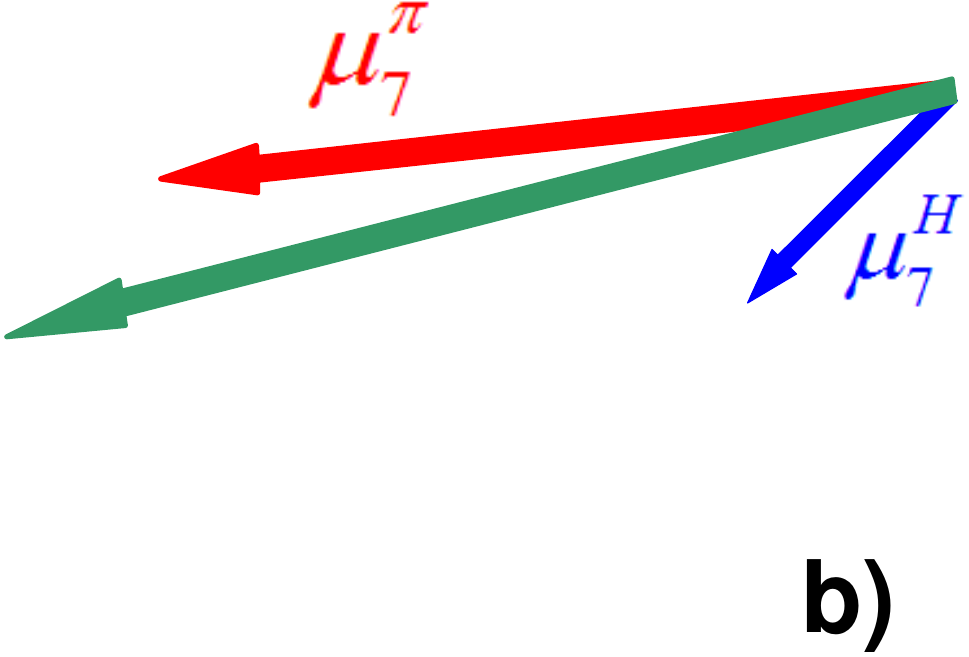}
	\includegraphics[angle=0,width=0.08\textwidth]{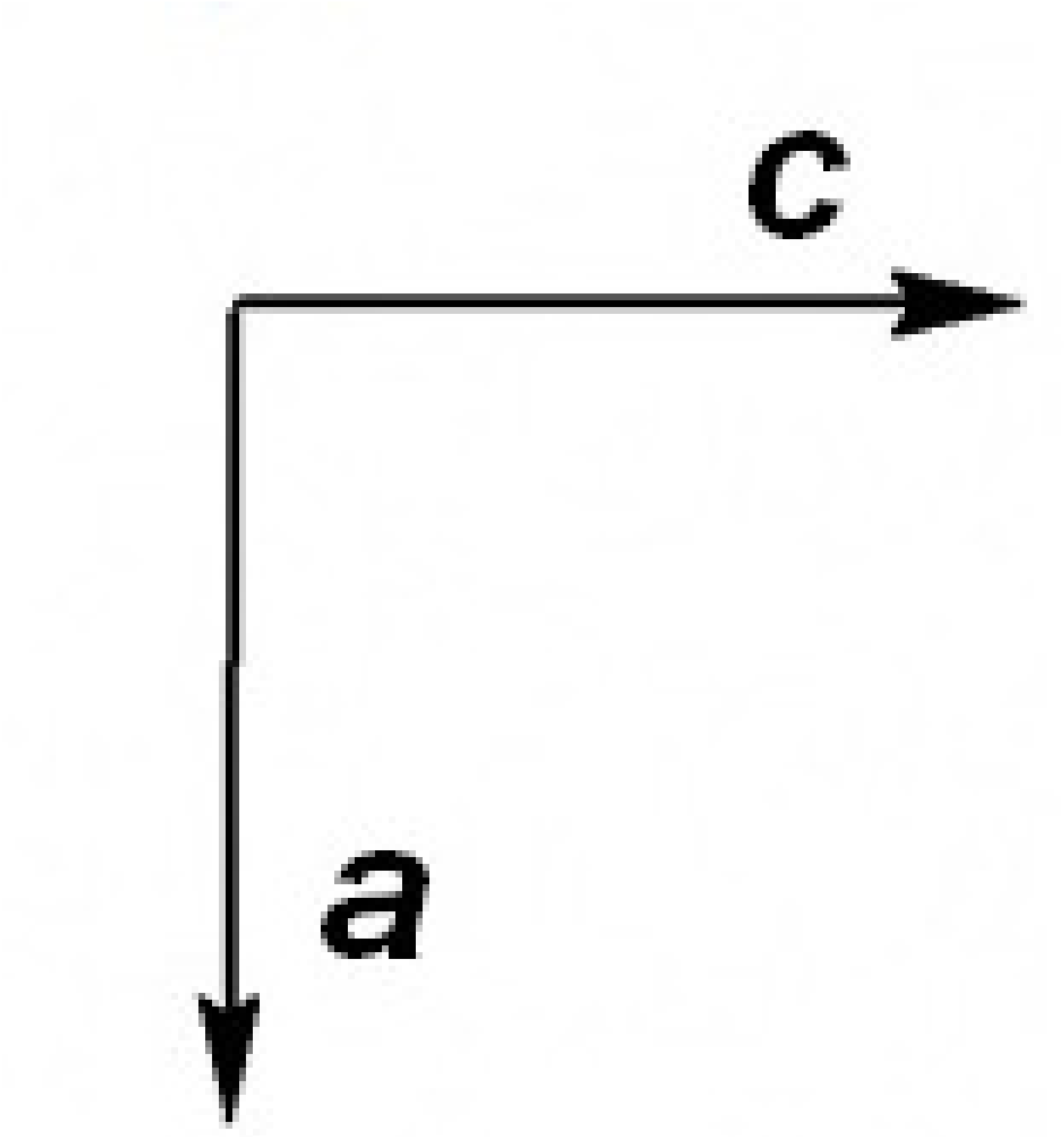}
\end{tabular}}
	\caption{(Colour online) Dipole moments of the lateral (a)  and single-ionized (b) proton configurations (configurations 1 and 7 of table~\ref{configurations_table}, respectively), as deduced from the results of \cite{horiuchi:18}.  ${\bm{\mu}}^H_1=(2\mu^H,0,0)$,
		$\bm{\mu}^\pi_1=(2\mu^\pi_\parallel,0,-2\mu^\pi_\perp)$; ${\bm{\mu}}^H_7=(\mu^H,0,-\mu^H)$,
		$\bm{\mu}^\pi_7=(\mu^\pi_\parallel-\mu^\pi_\perp,0,-\mu^\pi_\parallel-\mu^\pi_\perp)$.  Directions of the dipole moments associated with protons and with electrons are shown by blue and red arrows, respectively; the green arrows are the total dipole moment of the configuration; the vector lengths are nominal. $\varphi_0$ is the angle between the total dipole moment of configuration 1 and the $-c$ axis.
} \label{configurations_fig1}
\end{figure}

Our modelling of dipole moments of the H$_2$C$_4$O$_4$ groups 
is based on the results of the Berry phase calculations \cite{horiuchi:18}, which have shown that the sublattice polarization in this crystal is formed by the electronic contributions of switchable $\pi$-bond dipoles and directly by displacements of protons along the hydrogen bonds. Positions of the $\pi$-bonds are determined by the proton configurations around the given C$_4$O$_4$ group: in the ground state lateral configurations, the bond is formed between the two neighboring carbons, near  which protons sit on the hydrogen bonds (see figure~\ref{configurations_fig1}a), and also between the carbons and the adjacent oxygens, next to which there is no proton (meaning that the protons on these bonds sit in the minima close to the neighboring C$_4$O$_4$ groups). For the depicted lateral proton configuration, the  vector of the proton contribution to the dipole moment goes along the $a$ axis, while the total dipole moment is at the angle $\piup/2-\varphi_0$ to this axis. The three other lateral  configurations and their dipole moments can be obtained from the scheme of figure~\ref{configurations_fig1}a by rotation by a multiple of 90$^\circ$.

No Berry phase calculations for  the single-ionized configurations have been performed yet.  We assume~\cite{moina:21} that the angle between the protonic and electronic polarization vectors and the ratio of their absolute values are the same as in the case of the lateral configurations. For the lateral configuration 1 we have, according to figure~\ref{configurations_fig1}a,  ${\bm{\mu}}^H_1=(2\mu^H,0,0)$, $\bm{\mu}^\pi_1=(2\mu^\pi_\parallel,0,-2\mu^\pi_\perp)$. For 
configuration 7, the proton dipole moment is directed at 45$^\circ$ to the crystallographic axes, ${\bm{\mu}}^H_7=(\mu^H,0,-\mu^H)$, as it is easily determined from figure~\ref{configurations_fig1}b, and from the above assumptions it follows that $\bm{\mu}^\pi_7=(\mu^\pi_\parallel-\mu^\pi_\perp,0,-\mu^\pi_\parallel-\mu^\pi_\perp)$.  The dipole moments of the other single-ionized configurations can be obtained by rotation of the dipole moments of configuration 7. The dipole moment vectors and energies $\widetilde{{\cal E}_i}$ of all configurations in presence of the electric fields $E_1$ and $E_3$ are summarized in table~\ref{configurations_table}.

\renewcommand{\tabcolsep}{3.0pt}
\renewcommand{\arraystretch}{1.5}

\begin{table}[!t]
	\caption{Proton configurations and their energies $\widetilde{{\cal E}_i}$ in presence of the electric fields $E_1$ and $E_3$; $W_3=\mu^HE_1+\mu^\pi_{\parallel}E_1-\mu_\perp^\pi E_3$; 
	$W_1=-\mu^HE_3-\mu^\pi_{\parallel}E_3-\mu_\perp^\pi E_1$. Directions of the dipole moments associated with protons and with electrons are shown by blue and red arrows, respectively. }
	\vspace{0.5cm}
	\label{configurations_table}
	\begin{center}
		\small
		\begin{tabular}{c|c|c|c c c|c|c|c}
		\cline{1-4}  	\cline{6-9}
	$i$ &  &&$\widetilde{{\cal E}_i}$ & \hspace{7ex} & $i$ &  &&$\widetilde{{\cal E}_i}$\\
	\cline{1-4}  	\cline{6-9}
	1 & \includegraphics[height=1cm,width=1cm]{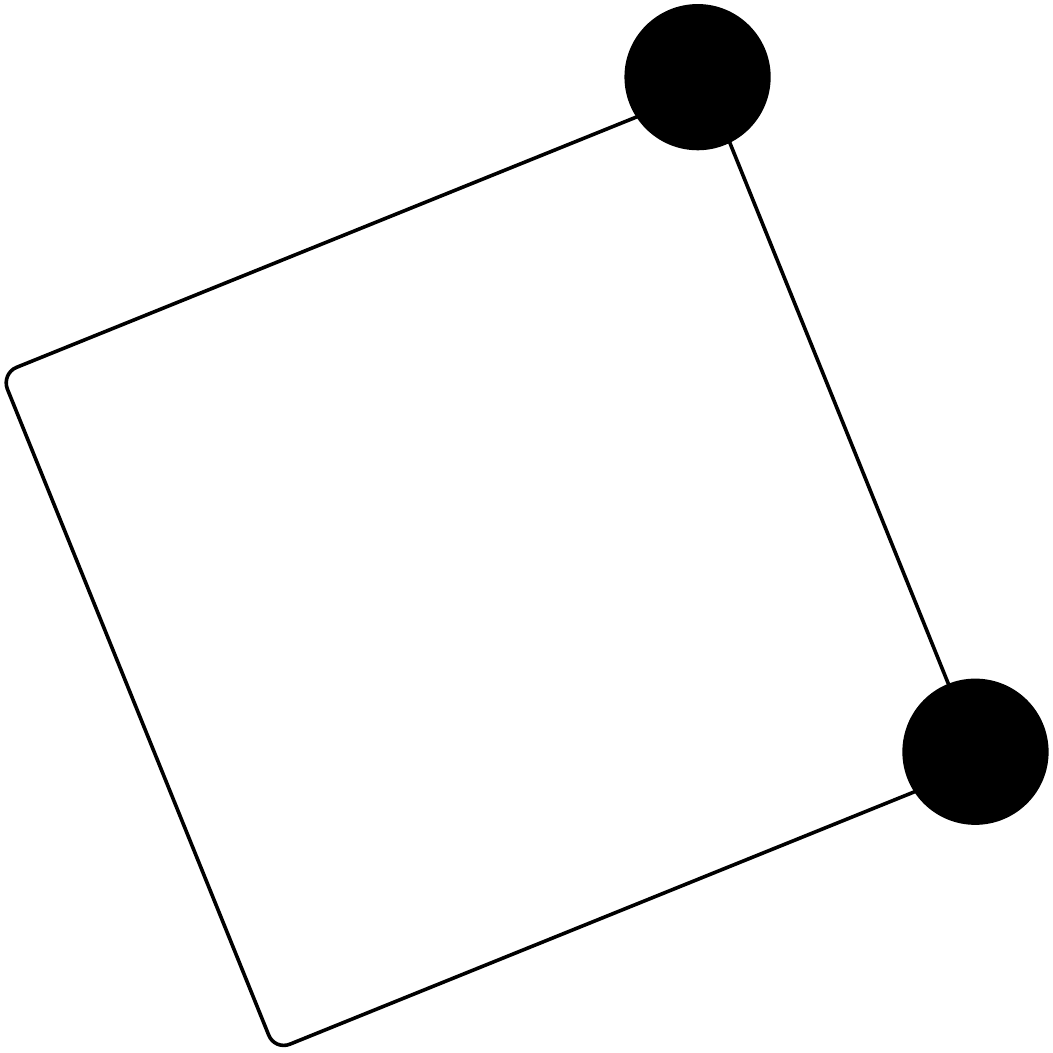} &   ~~~~\includegraphics[height=0.6cm,width=0.6cm]{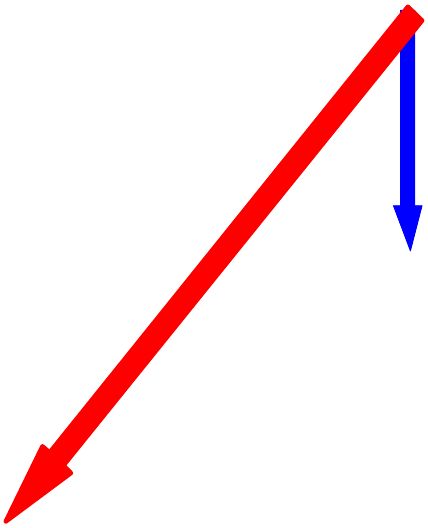} ~~~~ & ${\cal E}_a - 2W_3$ & & 	7 & \includegraphics[height=1cm,width=1cm]{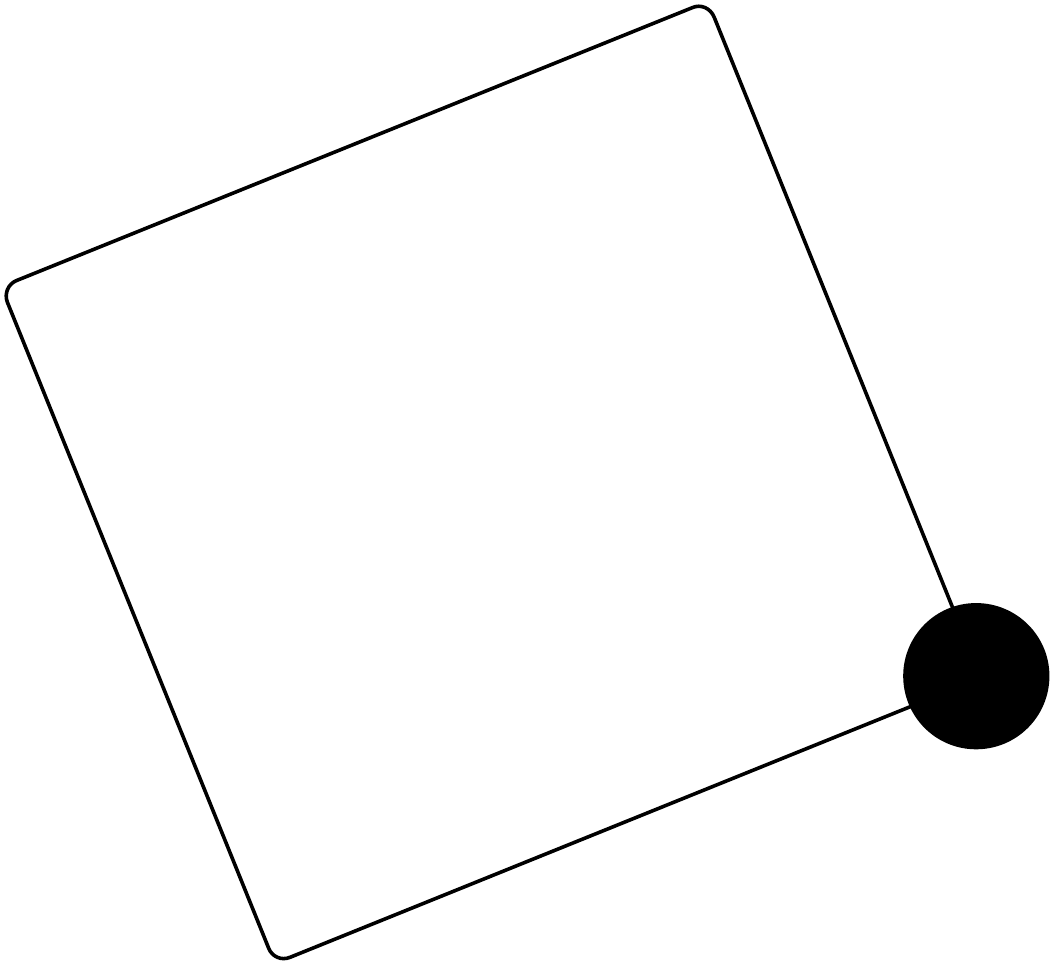}&    \includegraphics[angle=-45,origin=c,height=0.8cm,width=0.8cm]{arrows1} &${\cal E}_1-W_1-W_3$\\
	2 & \includegraphics[height=1cm,width=1cm]{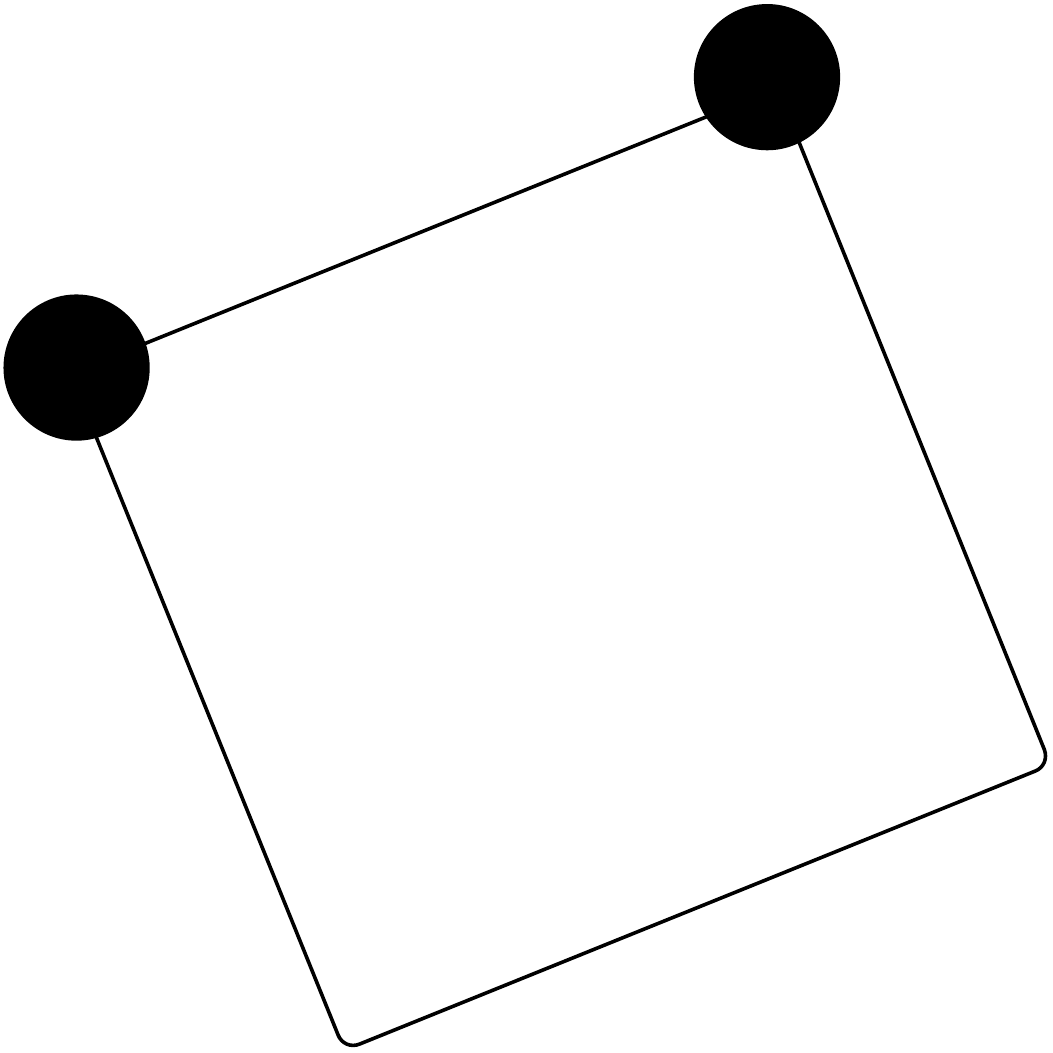}   
	&~~~~\includegraphics[angle=90,origin=c,height=0.6cm,width=0.6cm]{arrows1} ~~~~ &
	${\cal E}_a +2W_1$ & & 8 & \includegraphics[height=1cm,width=1cm]{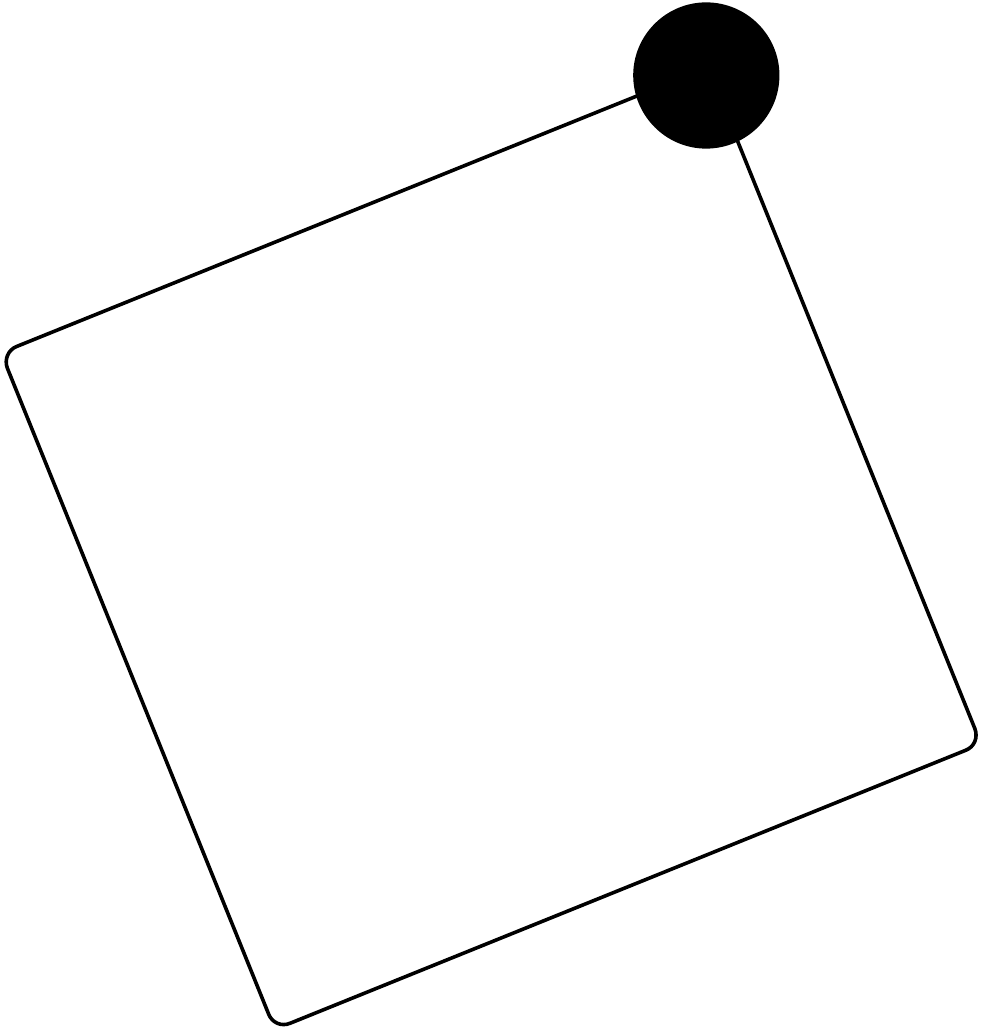}&   \includegraphics[angle=45,origin=c,totalheight=0.8cm,width=0.8cm]{arrows1} &${\cal E}_1+W_1-W_3$\\
	3 & \includegraphics[height=1cm,width=1cm]{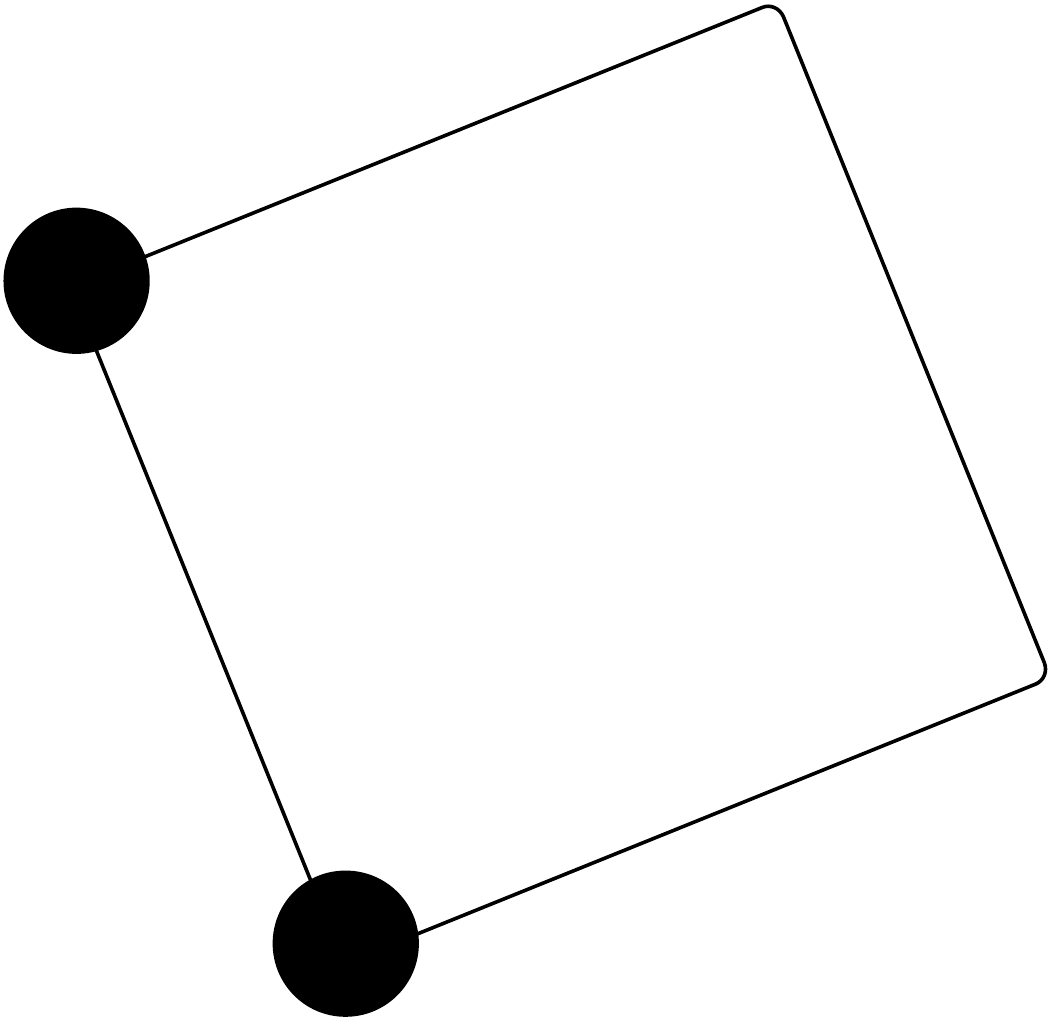} &   ~~~~\includegraphics[angle=180,origin=c,height=0.6cm,width=0.6cm]{arrows1} ~~~~ &
	${\cal E}_a + 2W_3$ && 	9 & \includegraphics[height=1cm,width=1cm]{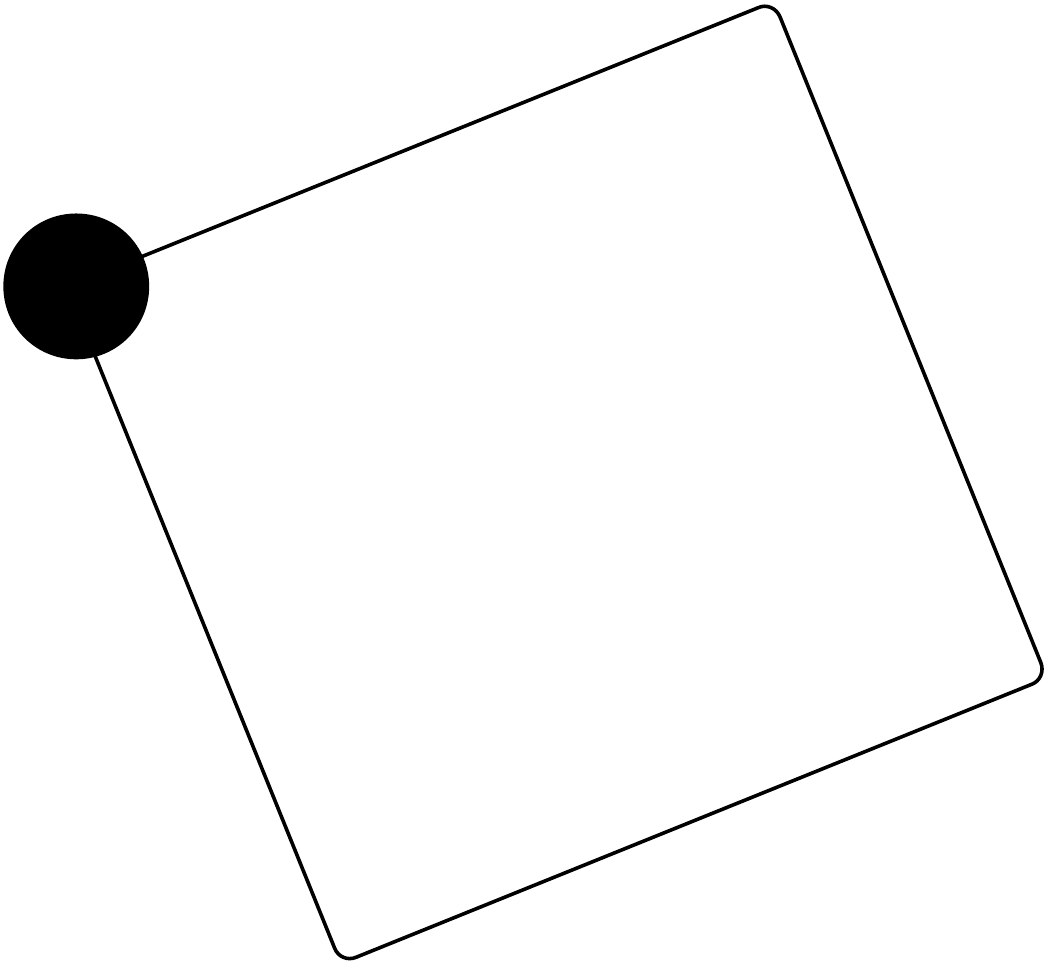} & 
	\includegraphics[angle=135,origin=c,height=0.8cm,width=0.8cm]{arrows1} & ${\cal E}_1+W_1+W_3$ \\
	4 & \includegraphics[height=1cm,width=1cm]{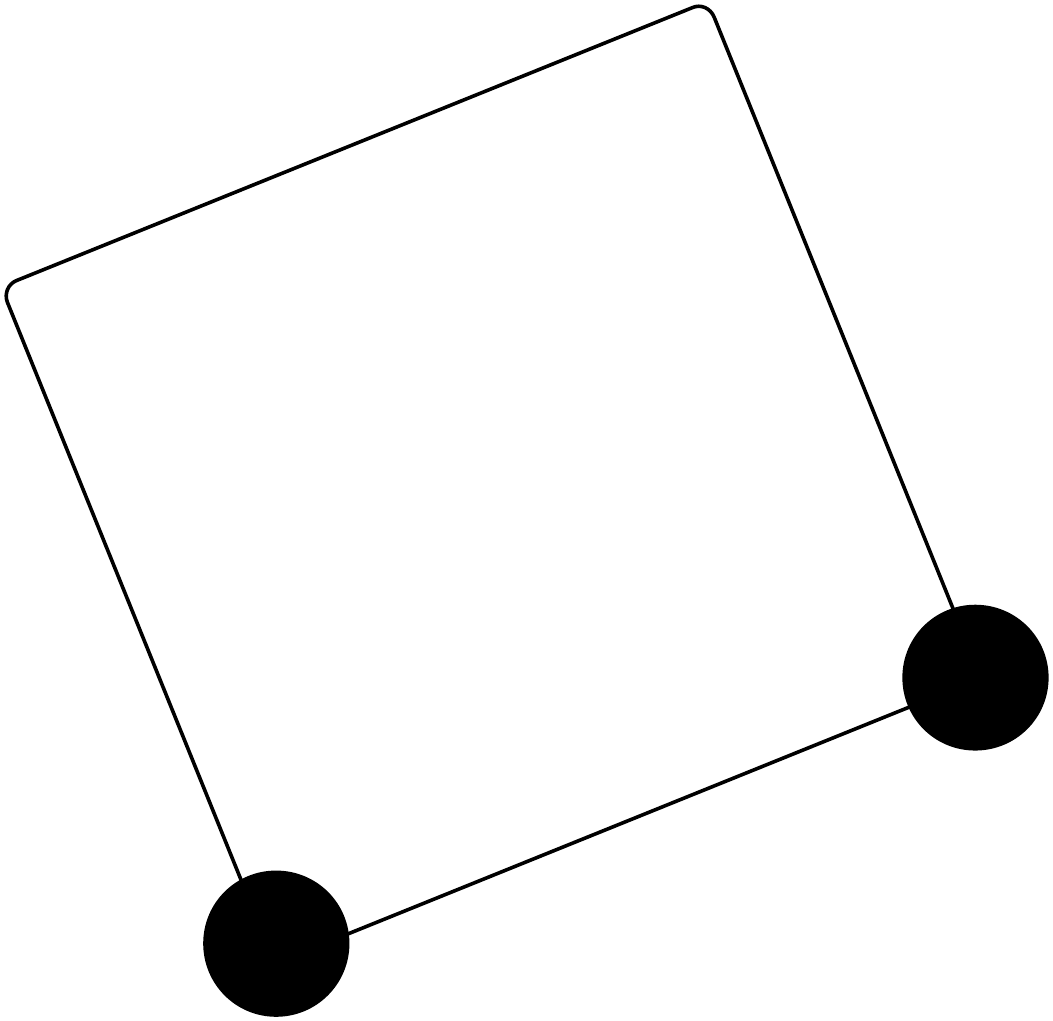} &  
	~~~~\includegraphics[angle=270,origin=c,height=0.6cm,width=0.6cm]{arrows1} ~~~~ &
		${\cal E}_a -2W_1$  && 10 & \includegraphics[height=1cm,width=1cm]{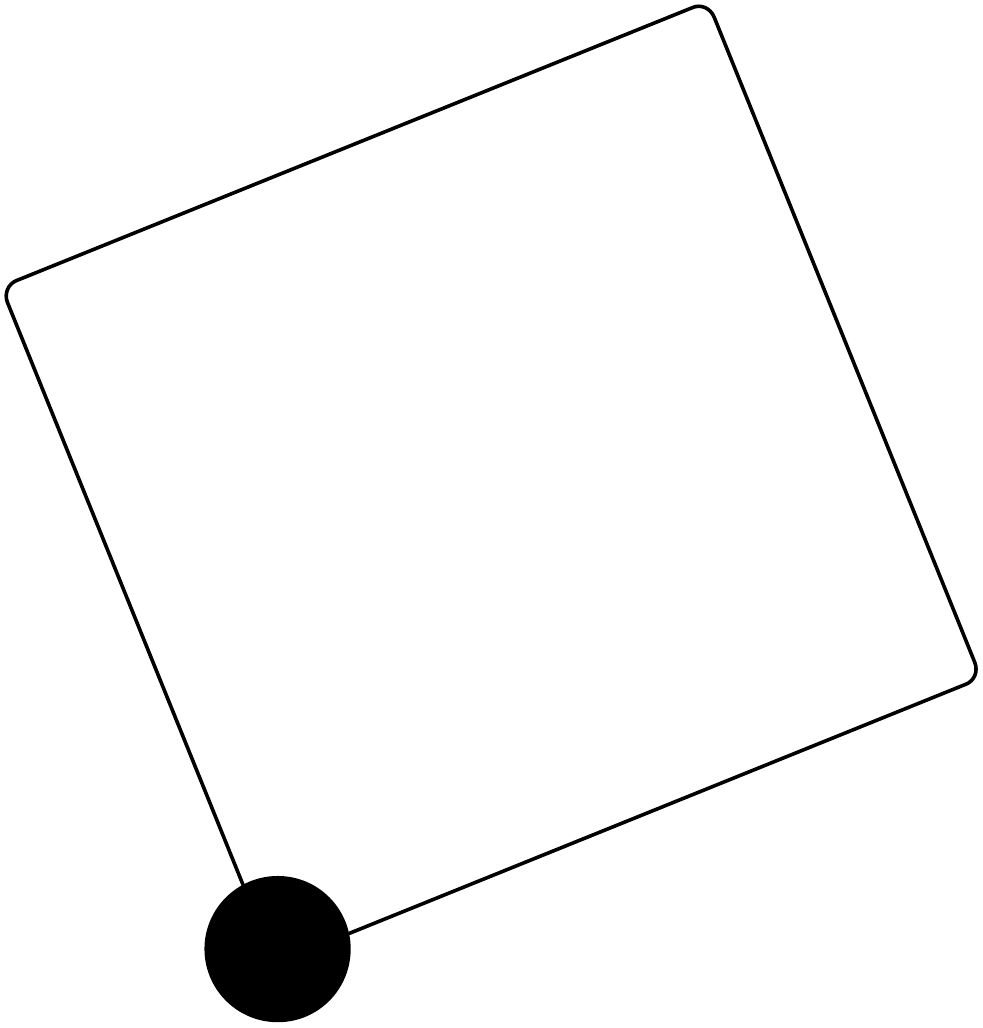}  &    \includegraphics[angle=225,origin=c,totalheight=0.8cm,width=0.8cm]{arrows1}  & ${\cal E}_1-W_1+W_3$ \\
\cline{1-4}
	5 & \includegraphics[height=1cm,width=1cm]{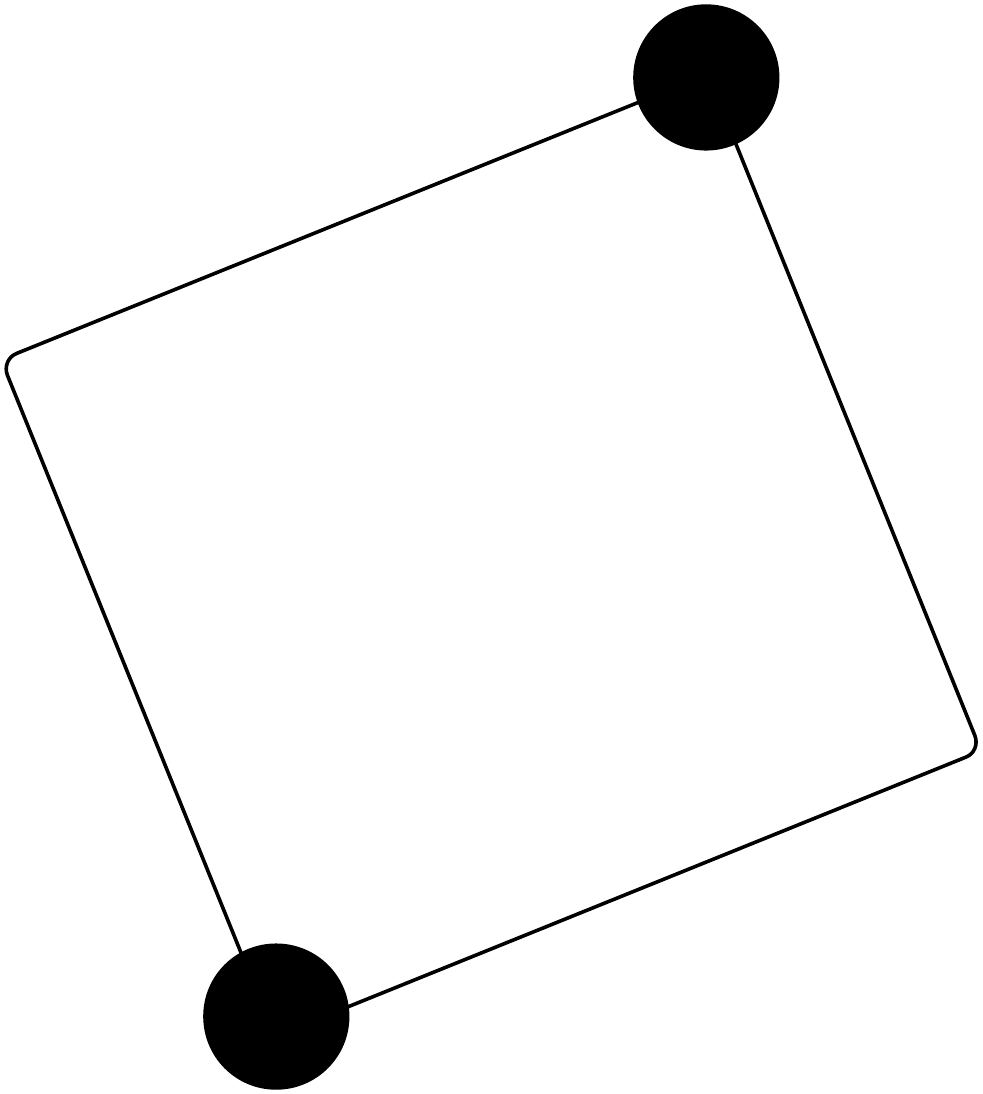} &   & ${\cal E}_s$ && 	11 & \includegraphics[height=1cm,width=1cm]{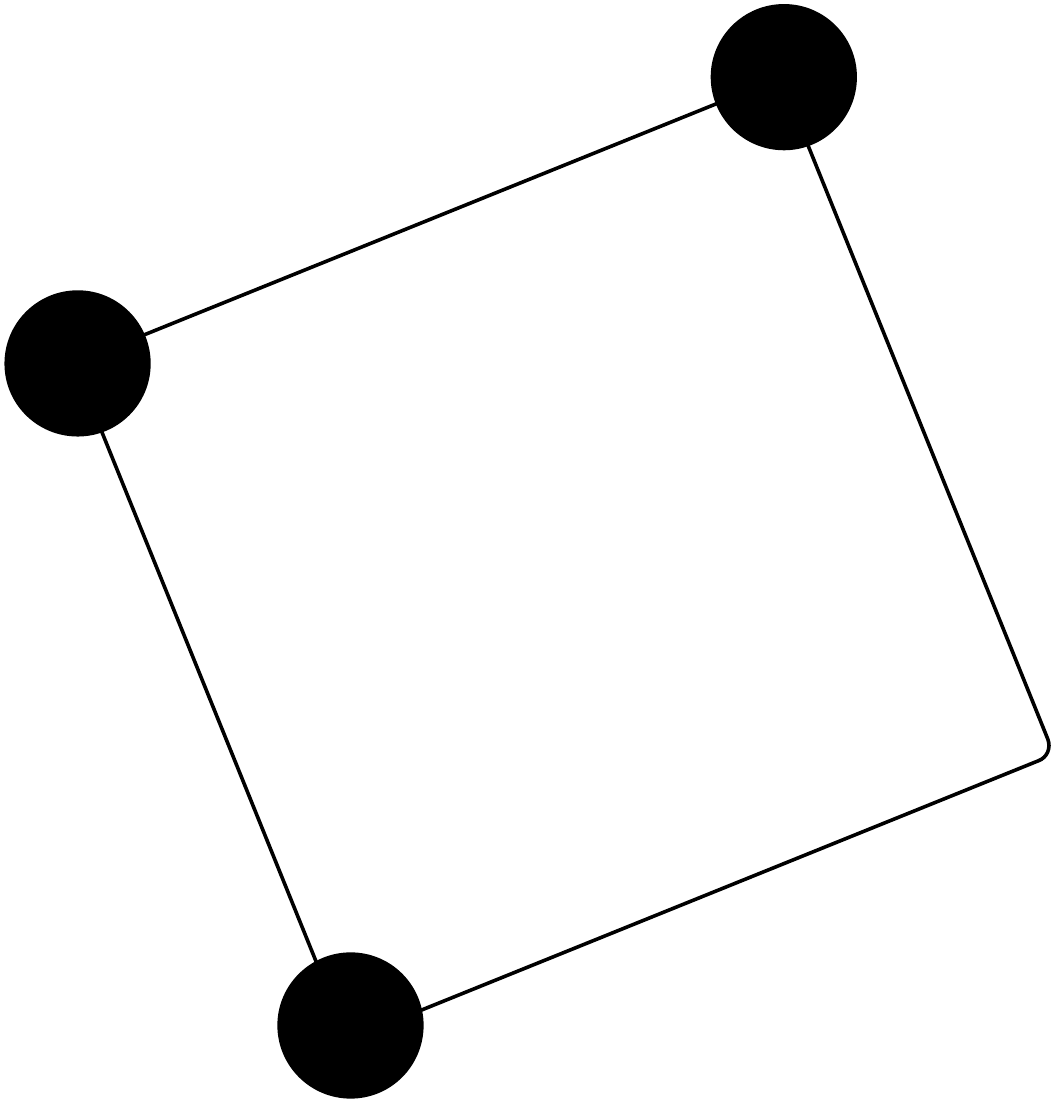}  &    \includegraphics[angle=135,origin=c,height=0.8cm,width=0.8cm]{arrows1} &  ${\cal E}_1+W_1+W_3$ \\
	6 & \includegraphics[height=1cm,width=1cm]{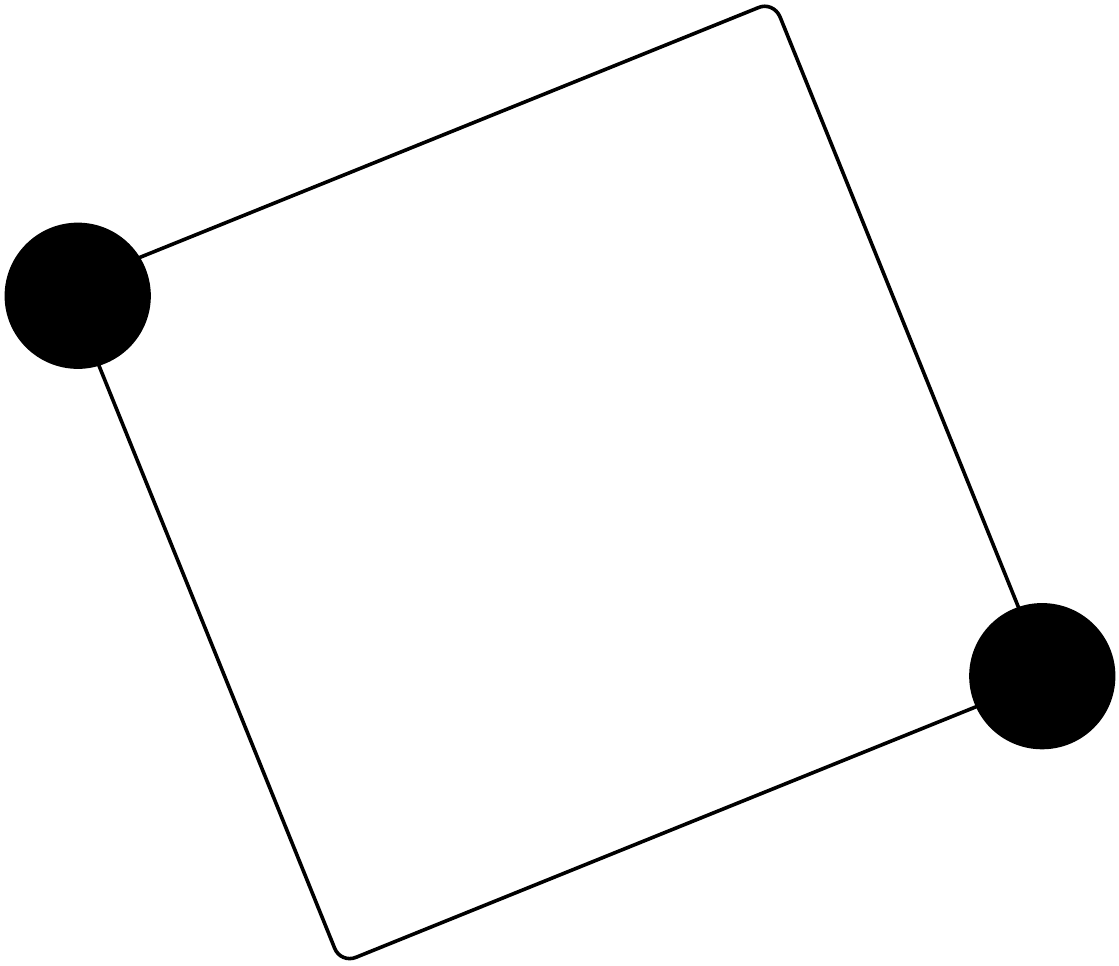} &   & ${\cal E}_s$ && 	12 & \includegraphics[height=1cm,width=1cm]{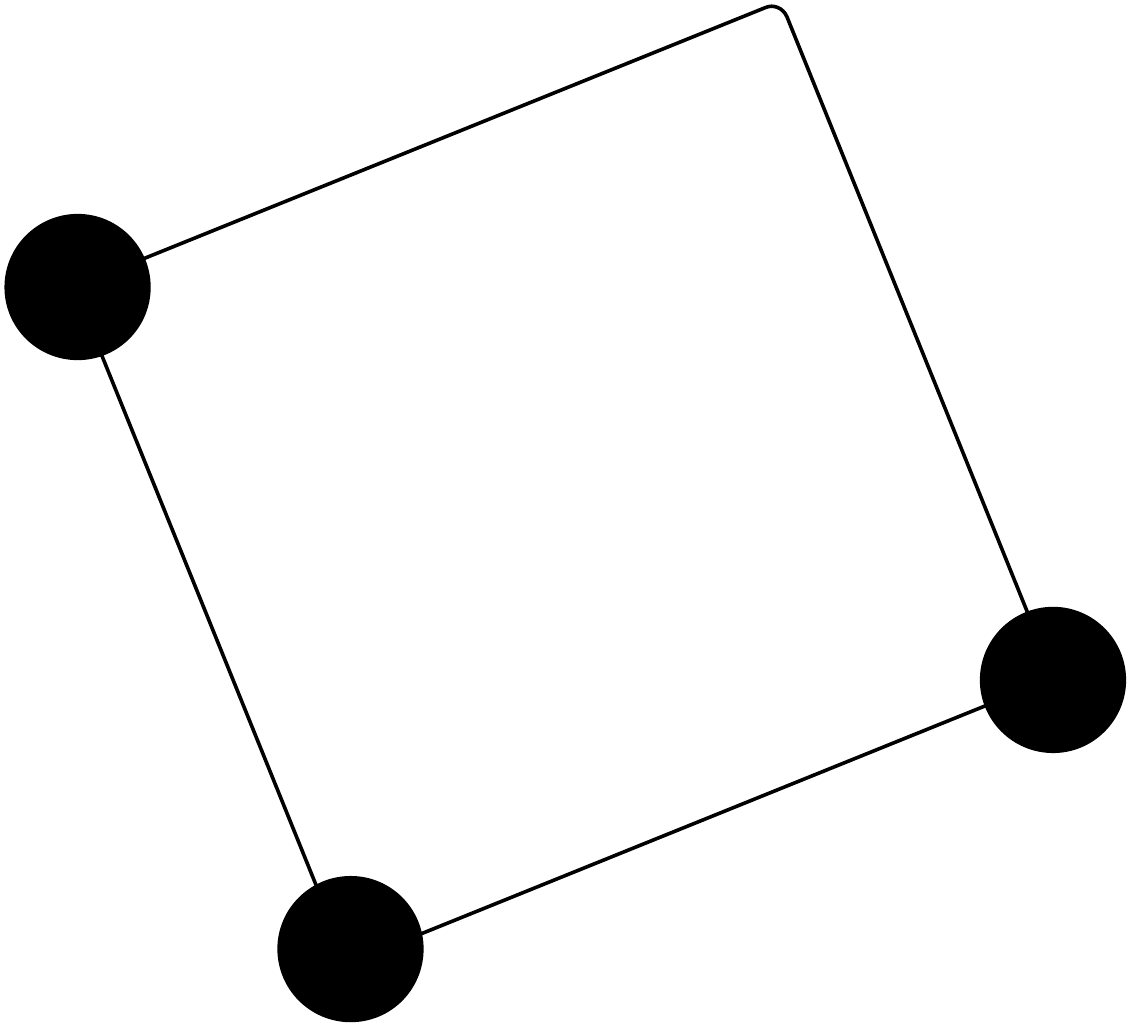} &    \includegraphics[angle=225,origin=c,totalheight=0.8cm,width=0.8cm]{arrows1} & ${\cal E}_1-W_1+W_3$ \\
\cline{1-4}
	15 & \includegraphics[height=1cm,width=1cm]{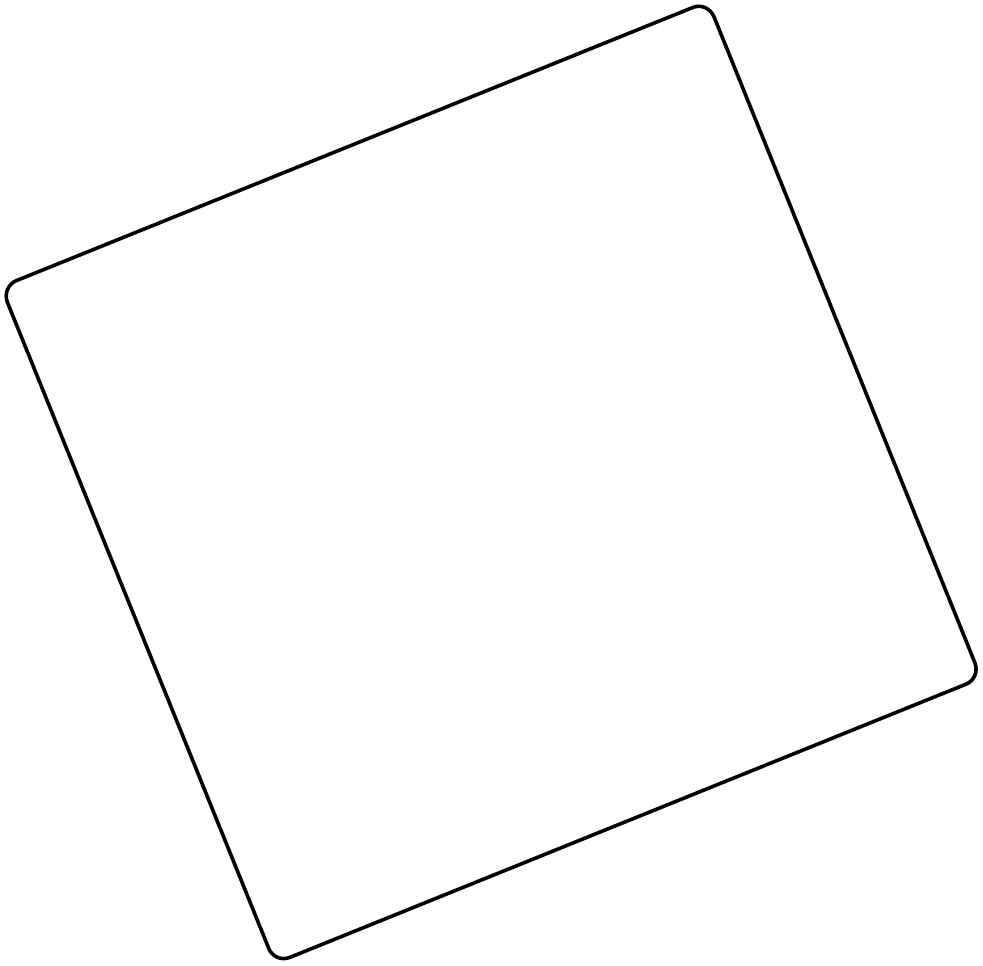}  && ${\cal E}_0$  && 	13 & \includegraphics[height=1cm,width=1cm]{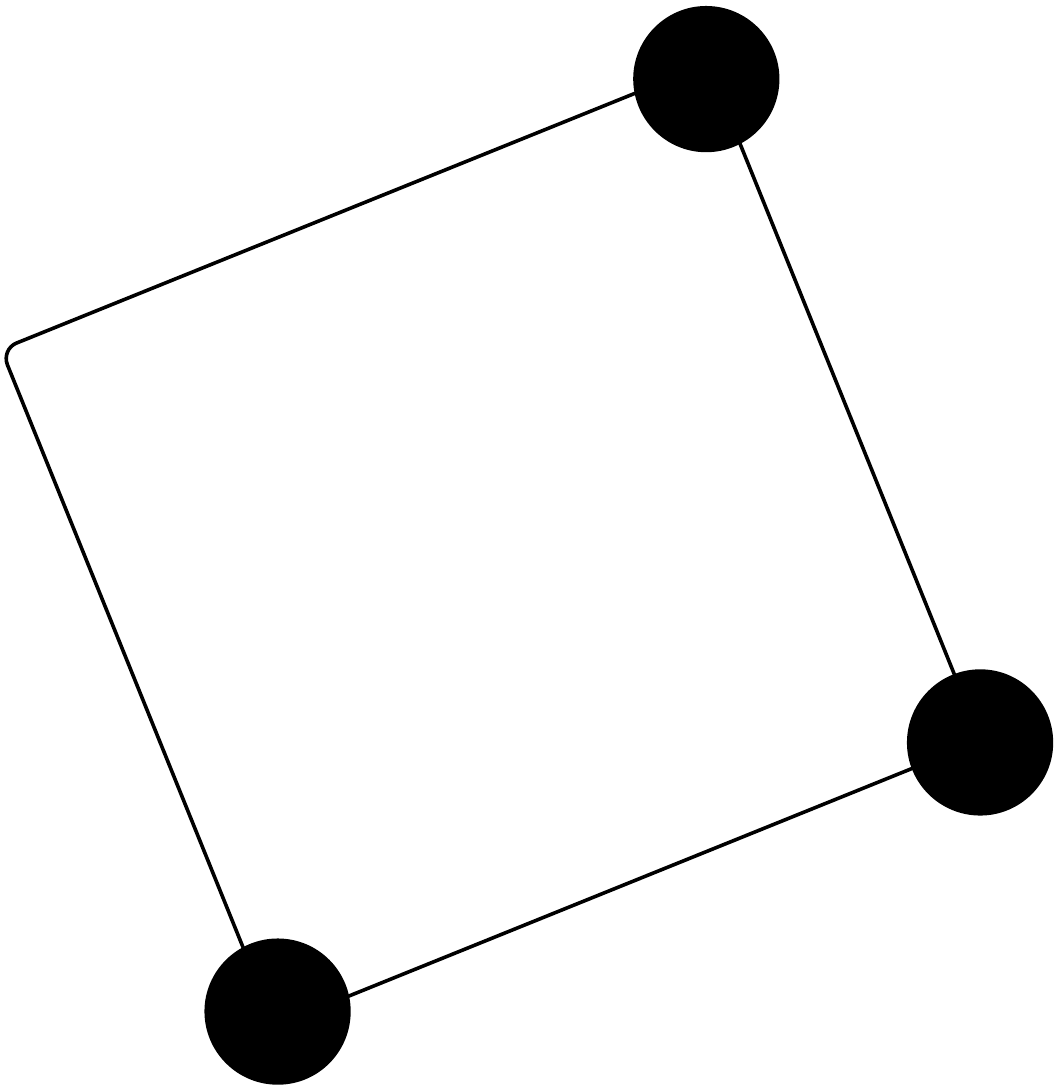}  &    \includegraphics[angle=-45,origin=c,height=0.8cm,width=0.8cm]{arrows1} & ${\cal E}_1-W_1-W_3$ \\
	16 & \includegraphics[height=1cm,width=1cm]{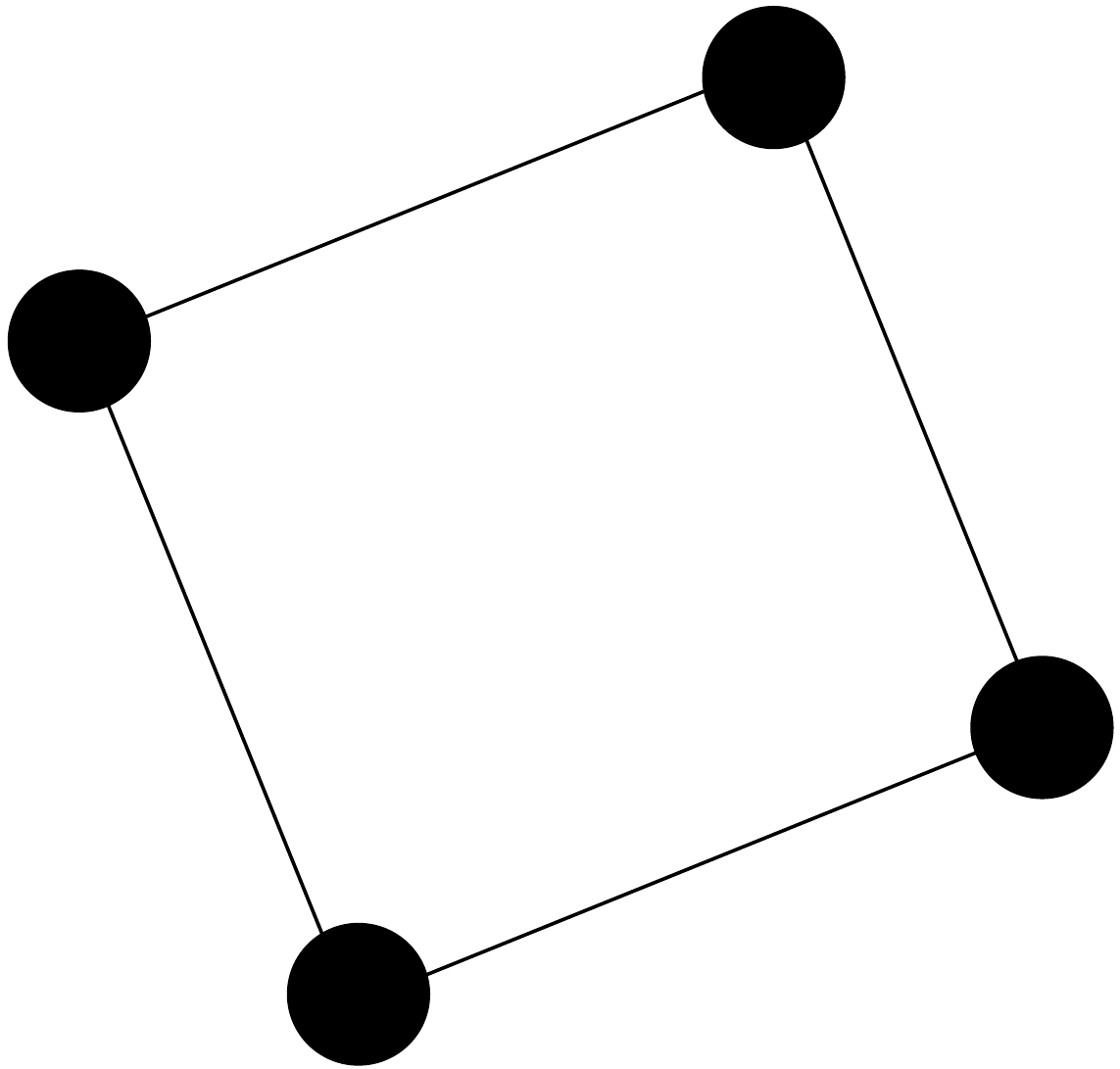} & & ${\cal E}_0$ && 14 & \includegraphics[height=1cm,width=1cm]{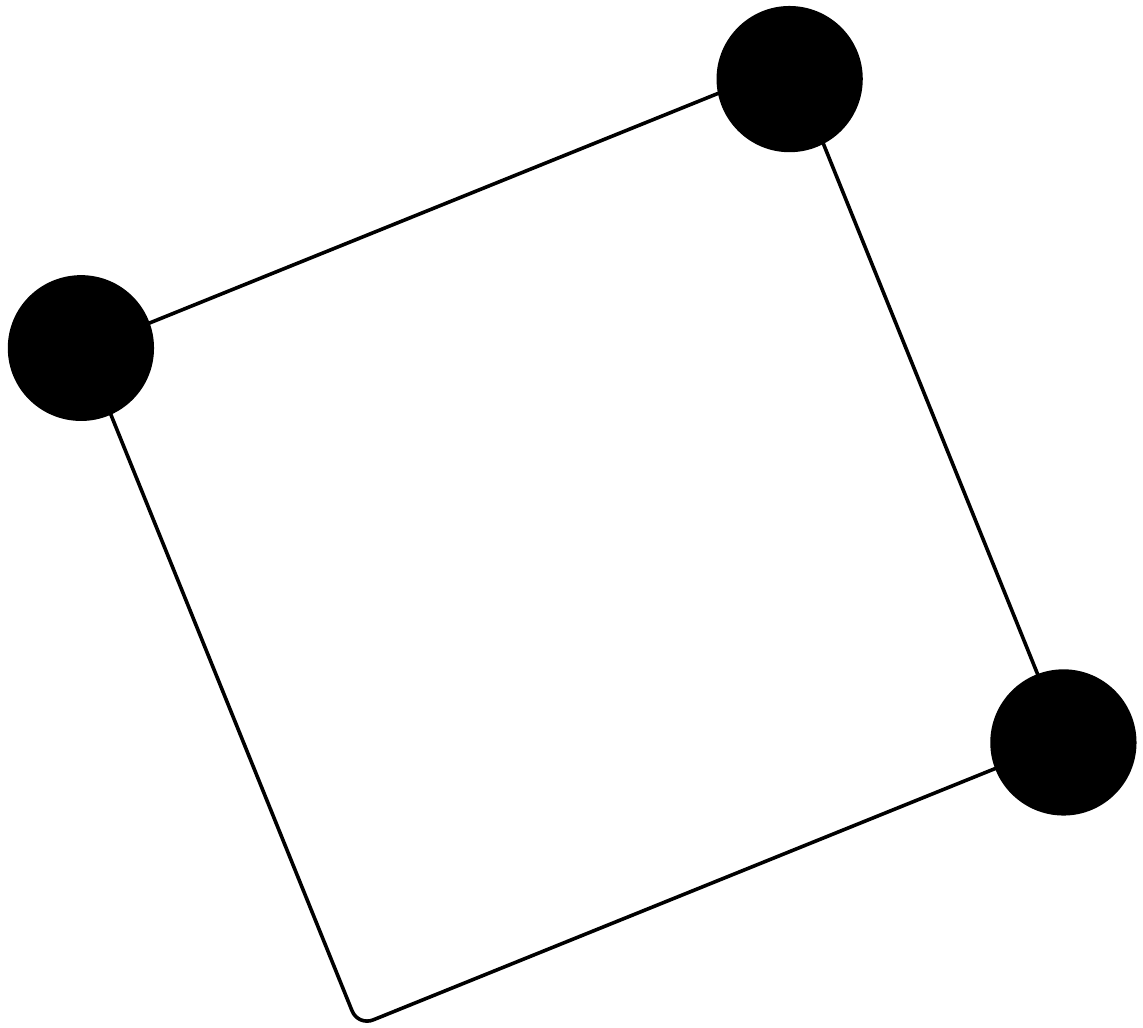}  &    \includegraphics[angle=45,origin=c,totalheight=0.8cm,width=0.8cm]{arrows1} & ${\cal E}_1+W_1-W_3$ \\
		\cline{1-4}  	\cline{6-9}
\end{tabular}
	\end{center}
\end{table}

To go from the representation of proton configuration energies to the pseudospin representation we use the standard procedure \cite{blinc:66,levitskii:04,moina:20,moina:21}, and the four-particle cluster approximation is employed for the obtained Hamiltonian.
The mean field approximation (MFA) is used for the long-range interlayer and intralayer interactions \cite{moina:20,moina:21}.

Eventually, the thermodynamic potential per one
 unit cell is obtained in the following form \cite{moina:21}
\begin{eqnarray}
&& g=U_{\textrm{seed}} -\frac1\beta[\ln D_+ +\ln D_-] -\frac1{2\beta}\left[\ln(1-\eta^2_{1+}) +\ln(1-\eta^2_{2+}) +
\ln(1-\eta^2_{1-})+
\ln(1-\eta^2_{2-})\right]\nonumber \\
&&\label{sqa:tpot} {}+ \nu\frac{(\eta_{1+}-\eta_{1-})^2}4+
 \nu\frac{(\eta_{2+}-\eta_{2-})^2}4+
\nu'\frac{(\eta_{1+}+\eta_{1-})^2}4+
\nu'\frac{(\eta_{2+}+\eta_{2-})^2}4,
\end{eqnarray}
where
\begin{align}
	 D_{\pm}&=a+\cosh(z_{1\pm}+z_{2\pm})+2b(\cosh z_{1\pm}+\cosh z_{2\pm})+\cosh(z_{1\pm}-z_{2\pm}),\nonumber \\
		 z_{f\pm}&=\frac12\ln\frac{1+\eta_{f\pm}}{1-\eta_{f\pm}}\pm\beta\nu\frac{\eta_{f+}-\eta_{f-}}{2} + \beta\nu'\frac{\eta_{f+}+\eta_{f-}}{2}
	+
	\beta\frac{\mu_{f1}E_1}{2}+\beta\frac{\mu_{f3}E_3}{2}, \nonumber\\
	 a&=\exp(-\beta\eps), \quad b=\exp(-\beta w). 
	\label{z}
\end{align}
The ``seed'' energy \cite{moina:20} 
\begin{equation}\label{key2}
U_{\textrm{seed}}=\frac {v}2\sum_{ij=1}^3c_{ij}^{(0)}u_iu_j-v\sum_{ij=1}^3c_{ij}^{(0)}\alpha_i^{(0)}\left(T-T_i^0\right)u_j
\end{equation}
corresponds to the deformable host lattice of heavy ions that forms the double-well potentials for the motion of protons. It
contains elastic and thermal expansion contributions associated with uniform diagonal lattice strains $u_1$, $u_2$, and $u_3$; $c_{ij}^{(0)}$ and $\alpha_i^{(0)}$ are the ``seed'' elastic constants and thermal expansion coefficients; $T_i^0$ determine the reference point of the thermal expansion of the crystal, which can be chosen arbitrarily; $v$ is the unit cell volume at this reference point. Inclusion of $U_{\textrm{seed}}$ into the Hamiltonian allows us \cite{moina:20} to correctly describe  the regular thermal expansion of squaric acid in the paraelectric phase and the hydrostatic pressure dependence of the lattice constants. 

The dipole moments $\mu_{f1}$ and $\mu_{f3}$ are as follows
\be
\label{dipole1}
\mu_{11}=-\mu_{31}=\mu^H+\mu^\pi_\parallel-\mu^\pi_\perp, 
\quad \mu_{21}=-\mu_{41}=\mu^H+\mu^\pi_\parallel+\mu^\pi_\perp, 
\ee
and
\be
\label{dipole3}
\mu_{13}=-\mu_{33}=-\mu^H-\mu^\pi_\parallel-\mu^\pi_\perp, 
\quad \mu_{23}=-\mu_{43}=\mu^H+\mu^\pi_\parallel-\mu^\pi_\perp.
\ee

The short-range Slater-Takagi type energy parameters
\[
\eps={\cal E}_s-{\cal E}_a, \quad w={\cal E}_1-{\cal E}_a, \quad w_1={\cal E}_0-{\cal E}_a,
\] according to the model \cite{moina:20}, are considered to be quadratic functions of the H-site distance $\delta$, and it has been assumed in equation~(\ref{z}) that $w_1\to\infty$.  In its turn, the distance $\delta$ is  taken to vary according to its experimentally observed \cite{semmingsen:95} above the transition  linear temperature dependence
\begin{equation}
\label{delta-model}
\delta=\delta_0[1+\delta_T(T-T_{\textrm{N0}})],
\end{equation}
where $T_{\textrm{N0}}$ is the transition temperature at zero electric field. It yields
\begin{equation}
\label{kdp-Slater}
\eps=\eps'_0[1+\delta_T(T-T_{\textrm{N0}})]^2, \quad w=w_0[1+\delta_T(T-T_{\textrm{N0}})]^2.
\end{equation}


The parameters $\nu$ and $\nu'$ from (\ref{sqa:tpot}) are linear combinations of the Fourier transforms, taken at the center of the Brillouin zone and at ${\bf k}_2=(0,b_2,0)$, of the eigenvalues of the matrices of the long-range (dipole-dipole and lattice mediated) interactions. Here $b_2$ is the basis vector of the reciprocal lattice~\cite{moina:20,moina:21}. For them, both the dependence of the dipole moments on $\delta$ and the variation
of the interactions with the changes in the equilibrium distances between protons (dipoles) are taken into account \cite{moina:20,moina:21}
\begin{equation}
\nu=\nu_0[1+\delta_T(T-T_{\textrm{N0}})]^2+\sum_{i=1}^3\psi_iu_i,
\label{kdp-long} \quad \nu'=\nu'_0[1+\delta_T(T-T_{\textrm{N0}})]^2+\sum_{i=1}^3\psi_i'u_i.
\end{equation}


The  plus and minus sublattice polarization vectors are 
\begin{equation}
\label{ppm}
\mathbf{P}_\pm=(P_{1\pm},0,{P_{3\pm}}), 
\end{equation}
where \cite{moina:21} 
\begin{eqnarray}
&& P_{1\pm} = 
\frac{\mu^H+\mu^\pi_\parallel}{2v}
(\eta_{1\pm}+\eta_{2\pm}) + \frac{\mu_\perp^\pi}{2v}(\eta_{2\pm}-\eta_{1\pm}),\nonumber \\
&& P_{3\pm}=
\frac{\mu^H+\mu^\pi_\parallel}{2v}
(\eta_{2\pm}-\eta_{1\pm}) - \frac{\mu_\perp^\pi}{2v}(\eta_{2\pm}+\eta_{1\pm}).
\label{net}
\end{eqnarray}

The order parameters $\eta_{f\pm}$ are found by numerical minimization of the thermodynamic potential (\ref{sqa:tpot}). At the same time, the strains $u_i$ are determined from the conditions that $\partial g/\partial u_i=0$ (see \cite{moina:20,moina:21}).

The noncollinearity angle $\theta$ is defined  as the difference of the angles between the current plus and minus sublattice polarization vectors (\ref{ppm}) at the given temperature and field and the polarization vector~$\mathbf{\bar P}_+$ of the fully ordered plus sublattice at low fields  \cite{moina:21}. The last vector is determined from equations~(\ref{ppm}), (\ref{net}) with the values of $\eta_{f+}=\bar\eta_{f+}$, corresponding to the ground state proton configuration of the plus sublattice. As the plus sublattice we take the one, the polarization in which is not switched by high positive fields. Since $\bar\eta_{f+}$ could only be either 1 or $-1$, we obtain
\begin{equation}
\label{theta}
\theta=\left|\arccos\frac{\bar\eta_{1+}\eta_{1+}+\bar\eta_{2+}\eta_{2+}}{\sqrt{2(\eta_{1+}^2+\eta_{2+}^2)}}-\arccos\frac{\bar\eta_{1+}\eta_{1-}+\bar\eta_{2+}\eta_{2-}}{\sqrt{2(\eta_{1-}^2+\eta_{2-}^2)}}\right|.
\end{equation}

\section{Calculations}
\label{calculations}
\subsection{Fitting procedure}
The developed model allows us to explore the influence of the electric field of any arbitrary orientation within the $ac$ plane.  The field $E$ in question with the components $(E_1,E_3)$ is directed at the angle $\varphi_E$ to the axis $a$; $\varphi_E$ is assumed to increase at a counterclockwise rotation. At $E_1>0$, $\varphi_E=\arctan E_3/E_1$. 



The values of all model parameters were determined earlier \cite{moina:20,moina:21}. In particular, they were required to provide the best fit to the experimental temperature curves  of
the order parameter at ambient pressure, the temperature and hydrostatic pressure dependences of the lattice strains  $u_i$, and the pressure dependence of  the transition temperature $T_{\textrm N}$.  The adopted values of the model parameters are given in table~\ref{tbl1}. A technical error, unfortunately, occurred in~\cite{moina:21}; as a result, the incorrect values of 
$\mu^H+\mu_\parallel^\pi$ and $\mu_\perp^\pi$ were given therein. Here, the correct numbers are presented.

From figure~\ref{configurations_fig1} and the values of $\mu^H+\mu_\parallel^\pi$ and $\mu_\perp^\pi$ from table~\ref{tbl1} it follows that the angle $\varphi_0$ between
the vector of the total dipole moment of configuration 1 and the $-c$ axis (or of configuration 2 and the $a$ axis) is about $56^\circ$
for this set of the model parameters. Note that from the Berry phase calculations \cite{horiuchi:18} it follows that the respective angle is about $50^\circ$.

The parameter $\nu'_0$ and the dipole moments  $\mu^H+\mu_\parallel^\pi$ and $\mu_\perp^\pi$ were determined by fitting the calculated curve of the dielectric permittivity $\eps_{11}$  at zero external bias field to the experimental points of~\cite{horiuchi:18},  while simultaneously trying to get as good as possible agreement with the experiment for the 
values of the switching fields, corresponding to the first 90$^\circ$ rotation of the sublattice polarization \cite{moina:21}. Unfortunately, the switching fields calculated with these parameters  are still considerably higher than the measured ones. The theoretical values of total polarization in the fully ordered NC90 and FE phases, induced by the field $E_1$, are also much higher \cite{moina:21} than predicted by the Berry phase calculations or measured experimentally (in the NC90 phase) \cite{horiuchi:18}, and so is the value of the saturated spontaneous sublattice polarization in the AFE phase.

Alternatively, $\mu^H+\mu_\parallel^\pi$ and $\mu_\perp^\pi$ can be chosen by fitting the theory to the values of polarization, found from the Berry phase calculations \cite{horiuchi:18}. However, it will essentially spoil the agreement with the experiment for the permittivity and even further increase the difference between the theory and experiment for the switching fields. It appears that all these three quantities (permittivity, polarization, switching fields) cannot be well described by the present model with the same set of the fitting parameters. This controversy can possibly be addressed by taking into account the non-linear terms in the Hamiltonian, e.g., of the type $\mu'(\sum \sigma)^2\sigma E$, which is equivalent to the terms like $EP^3$ in the phenomenological approach, similarly to how it was proposed for a description of electric field effects in the KH$_2$PO$_4$ family crystals~\cite{vdovych:14,vdovych:16}.

\begin{table}[htb]
	\caption{The adopted values of the model parameters, taken from \cite{moina:20,moina:21}. }
	\label{tbl1}
	\begin{center}
		\renewcommand{\arraystretch}{0}
		\begin{tabular}{ccccccc|cccc}
			\hline
			$\eps'_{0}/k_{\textbf{B}}$ & $w_0/k_{\textbf{B}}$ &  $\nu_0/k_{\textbf{B}}$ &  $\nu'_0/k_{\textbf{B}}$ & $\psi_1/k_{\textbf{B}}$ & $\psi_2/k_{\textbf{B}}$ & $\psi_3/k_{\textbf{B}}$ &		$c_{11}^{0}$&  $c_{12}^{0}$ & $c_{13}^{0}$ & $c_{22}^{0}$  \\
			\multicolumn{7}{c|}{(K)}  &			\multicolumn{4}{c}{($10^{10}$ N/m$^{2}$)}  \\
			\hline\\
			395 &  1100 & 79.9 & $-50$ &  $-518$
			& 445 & 1096 &	6.5 & 2.3 &$-3.1$ & $2.38-0.02T$ \strut 	\\ 	\hline
		\end{tabular}
		\bigskip
		
		\begin{tabular}{ccc|c|cc}
			\hline
			$\alpha_1^0$ & $\alpha_2^0$ &$\delta_T$  & $v$ & $\mu^H+\mu^\pi_\parallel$ & $\mu^\pi_\perp$  \\
			\multicolumn{3}{c|}{($10^{-5}$ K$^{-1}$)}  & ($10^{-28}$m$^3$) 
			& \multicolumn{2}{c}{($10^{-29}$~Cm)} \\
			\hline
			1.2 & 13.0 & 20 & 2.0 &3.16 &2.12  \strut 	\\ 	\hline
		\end{tabular}

		\vspace{1ex}
		\renewcommand{\arraystretch}{1}
	\end{center}
\end{table}

Moreover, as extensively discussed  in~\cite{moina:21}, the adopted set of $\nu'_0$ and $\mu$ is not unique. The values of these parameters can be varied within certain limits, and other acceptable sets of the values can be found, which provide a similar agreement with the available experimental data  for the permittivity and switching fields. The calculated position of the upper field-induced transition line between the noncollinear ferrielectric and collinear ferroelectric phase is most sensitive to the values of these parameters. In view of the discussed uncertainty, this position should be treated with caution and as a qualitative theoretical prediction, rather than a quantitative estimate.

\subsection{Ground state phase diagram}
\label{groundstate}
As it has been already mentioned, our previous calculations \cite{moina:21} have revealed that the external field~$E_1$ at low temperatures induces two phase transitions in the squaric acid crystals, each associated with a sublattice polarization rotation by 90$^\circ$. The predicted sequence of the phases  with increasing field is as follows: nearly antiferroelectric with antiparallel sublattice polarizations (AFE*) --- noncollinear ferrielectric with perpendicular sublattice polarizations (NC90) --- collinear ferroelectric (FE). 

\begin{figure}[!t]
	\centerline{
\includegraphics[width=0.85\textwidth]{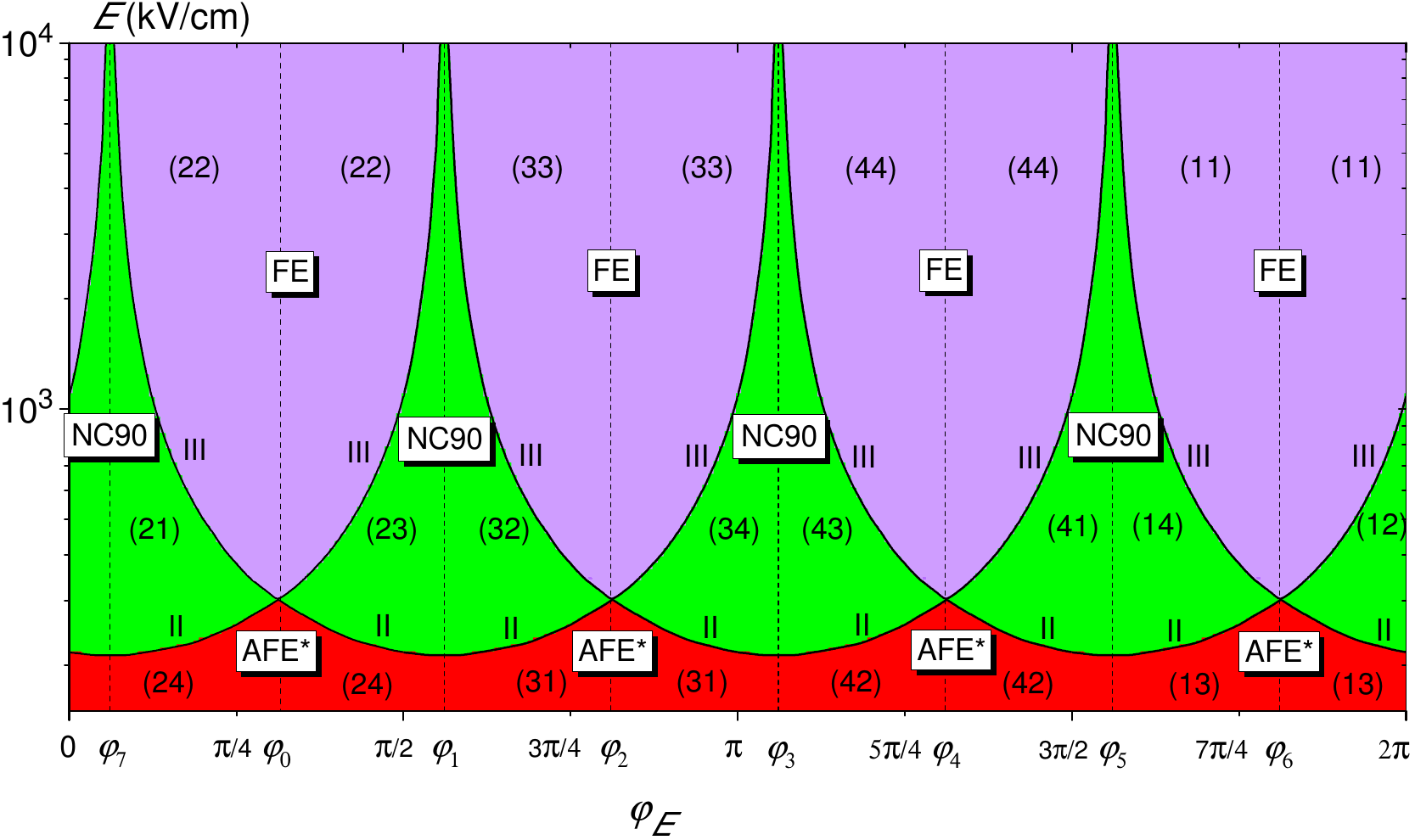}
}
	\caption{(Colour online) The $\varphi_E$--$E$ phase diagram of squaric acid at 250~K. The solid lines II and III are the lines of the first order phase transitions between the AFE* (the red region, $\theta\approx 180^\circ$) and NC90 (the green regions, $\theta\approx 90^\circ$) phases and between the NC90 and FE (the purple regions, $\theta=0$) phases, respectively. Vertical dashed lines are a guide to the eye. The numbers in parentheses indicate the ground state proton configurations in each phase in the plus and minus sublattices, respectively (table~\ref{configurations_table}). }
	\label{pdphi}
\end{figure}

In order to verify whether this behavior persists for
other directions of the electric field confined to the $ac$ plane, 
we construct the low-temperature $\varphi_E$--$E$ phase diagram of the squaric acid. The model symmetry in the absence of the field is pseudotetragonal, and the electric field applied within the $ac$ plane breaks this symmetry. The thermodynamic potential (\ref{sqa:tpot}) is invariant with respect to the 90$^\circ$ rotation around the $b$ axis, at which $E_1\to E_3$, $E_3\to -E_1$, $\varphi_E \to \varphi_E+\piup/2$, $\eta_{1\pm}\to\eta_{2\pm}$, $\eta_{2\pm}\to-\eta_{1\pm}$. 
It is thus expected that the $\varphi_E$--$E$ diagram should also be symmetric (in terms of the phases and transitions between them) with respect to the 90$^\circ$ rotation of the field $\varphi_E \to \varphi_E+\piup/2$. Figure~\ref{pdphi}, where the obtained $\varphi_E$--$E$ diagram is presented, proves that this is, indeed, the case. 

As one can see, the two-step polarization rotation and the AFE*--NC90--FE phase sequence are observed at all field orientations, except for eight special directions with $\varphi_E=\varphi_{n}$,  where $\varphi_n=\varphi_0+\piup n /4$, $n=0,...7$. The lines of the first order transitions between the AFE* and NC90 phases and between the NC90 and FE phases are denoted as II and III, following the notations adopted in \cite{moina:21}.

There are, obviously, two types of these special directions. The first type is at an even $n$, i.e., at $\varphi_E=\varphi_{2k}$, $k=0,1,2,3$. In this case, the electric field is collinear to the antiparallel sublattice polarizations of the AFE* phase (this collinearity is discussed in more detail later in this subsection), so the squaric acid becomes effectively a uniaxial antiferroelectric. The intermediate non-collinear phase NC90 disappears, and polarization switching is a traditional one-step process from AFE* to FE phase. The second type is at odd $n$, at $\varphi_E=\varphi_{2k+1}$, when the external electric field is applied at 45$^\circ$ to the sublattice polarizations of the AFE* phase. In this case, the non-collinear NC90 phase is the widest, with the upper transition field to the FE phase at lines III tending to infinity.

Experimentally, the low-temperature transition to the ferroelectric phase has not been detected yet  because of the dielectric breakdown of the samples \cite{horiuchi:18,Ol}. As it follows from the constructed $\varphi_E$--$E$ diagram, the field of this transition is the lowest when its direction is close to the axis of the sublattice polarization, $\varphi_E=\varphi_{2k}$. Measurements of the polarization switching for the fields oriented in the sector between $50^\circ$ and $60^\circ$ to the $a$ axis seem most promising in this respect.

The $\varphi_E$--$E$ phase diagram, in terms of the phases, phase transitions, and relative orientation of the sublattice polarization vectors described by the noncollinearity angle $\theta$, is  symmetric with respect to the field rotation by $90^\circ$. However, the absolute orientations of the sublattice polarization vectors for each $\varphi_E$ and for the corresponding $\varphi_E\pm\piup/2$ at the same field magnitudes are different. This is caused by the fact that different lateral proton configurations become the ground state ones in the same phase.

At low temperatures, all protons within a sublattice are in the same lateral configuration, one of the four possible (see table~\ref{configurations_table}). The ground state sublattice polarization vector, therefore, can be only oriented along one of the four directions, depending on which lateral configuration is realized in this sublattice. This configuration can be determined from the signs of the order parameters $\eta_{1\pm}$ and $\eta_{2\pm}$ at low temperatures, found by minimization of the thermodynamic potential (\ref{sqa:tpot}). Positive $\eta_{f\pm}=1$ or negative  $\eta_{f\pm}=-1$ mean that the proton on the $f$th bond in the plus/minus sublattice is localized at the site close to the considered C$_4$O$_4$ group or to the neighboring  C$_4$O$_4$ group, respectively. A change of the sign of the order parameter, therefore, means flipping of the proton at the respective bond. 
The obtained ground state configurations of the plus and minus sublattices in each region of the $\varphi_E$--$E$ phase diagram are indicated by the pairs of numbers in figures~\ref{pdphi} and \ref{circle}; the respective proton arrangements are visualized in figure~\ref{phasesE1} in the appendix.

\begin{figure}[!t]
	\centerline{
\includegraphics[width=0.5\textwidth]{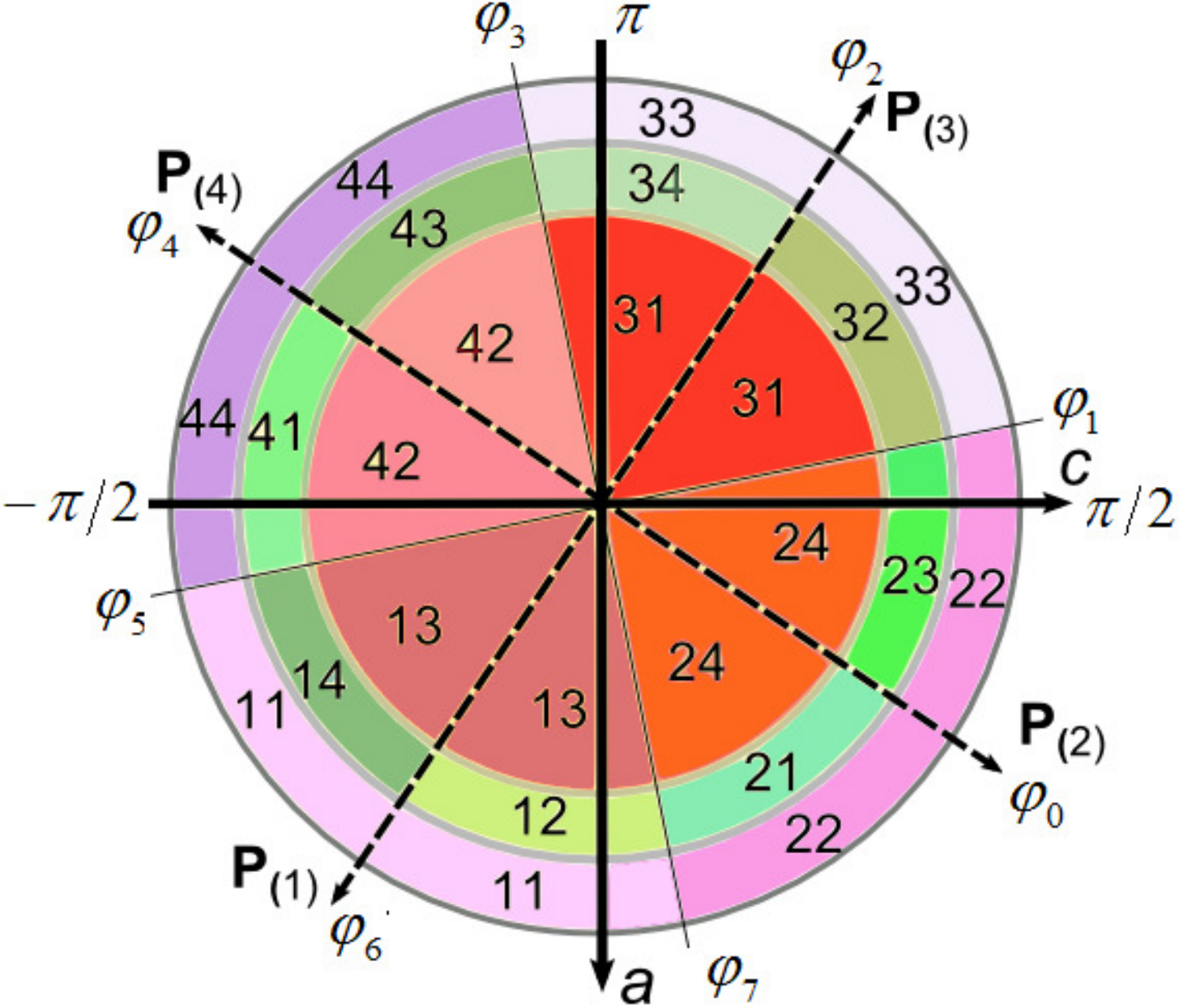}
	}
	\caption{(Colour online) The diagram of the ground state proton configurations vs angle $\varphi_E$.  The pairs of numbers in each segment or sector indicate the proton configurations of the plus and minus sublattices, respectively, in the AFE* (the inner circle, shades of red), NC90 (the intermediate segments, shades of green), FE (the outer segments, shades of purple) phases. Different shades correspond to different configurations. Bold black solid lines: crystallographic axes $a$ and $c$. Black dashed lines: axes of (pseudo)spontaneous polarization $\mathbf{P}_{(1)}$--$\mathbf{P}_{(3)}$ and $\mathbf{P}_{(2)}$--$\mathbf{P}_{(4)}$.  }
	\label{circle}
\end{figure}

The four possible directions of the low-temperature sublattice polarization vectors are shown in figure~\ref{circle} by the dotted lines as the axes $\mathbf{P}_{(1)}$--$\mathbf{P}_{(3)}$ and $\mathbf{P}_{(2)}$--$\mathbf{P}_{(4)}$.  They coincide with the directions of the total dipole moments of the lateral configurations 1--4 and oriented at the angles $\varphi_{2k}$ to the axis $a$. Each change of the configuration number $1\to 2\to3\to 4\to1$ corresponds to a counterclockwise rotation of the  polarization vector by $90^\circ$. Figure~\ref{phasesE1}  illustrates the polarization vectors for 
the two sublattices at full saturation in different regions of the $\varphi_E$--$E$ phase diagram.

The following conclusions can be drawn from the analysis of the 
ground state proton configurations at different field directions:

\begin{enumerate}
	\item
	The obvious one, that rotation of the field by $90^\circ$, $\varphi_E\to\varphi_E+\piup/2$, leads to a $90^\circ$ rotation of the sublattice polarization vectors in the same direction for all field magnitudes.
	
	\item
The ground state configurations are switched in a first order transition-like manner at $\varphi_{2k+1}$  in all phases and at $\varphi_{2k}$ in the NC90 phase only. 	Because of the switching at $\varphi_E=\varphi_{2k+1}$ and the resulting rotation of the polarization vectors by 90$^\circ$, which occur even at vanishingly small field magnitudes, in the AFE* phase the angle between field and the polarization axis never exceeds 45$^\circ$ and is zero at each $\varphi_{2k}$, where the field is collinear to the sublattice polarizations.
	
	\item 
	The minus sublattice configuration for the NC90 phase changes to the antiparallel one at each $\varphi_{2k}$. This has an obvious geometric explanation:  The sign of the tangential to the AFE* phase polarization vectors component of the electric field, which causes the 90$^\circ$ rotation of the polarization, is changed at $\varphi_{2k}$.
\end{enumerate}

The entire scheme of  figure~\ref{circle} stems from the low-field switching of the proton configurations in the AFE* phase at $\varphi_{2k+1}$. It creates the initial AFE* configuration of the system, starting from which the subsequent high-field rotation of the minus sublattice polarization and transitions to the NC90 and FE phases occur. Basically, this low-field switching is the change of the orientation of the (pseudo) spontaneous polarizations caused by the external field. In a perfect crystal, the paraelectric (tetragonal)~-- antiferroelectric (monoclinic) transition is a spontaneous symmetry breaking process in absence of external fields, and either the configurations 1 and 3 or 2 and 4 can become the ground state one below the transition. Similarly, either of the tetragonal axes $a_t$ or $c_t$ above the transition can become the $a$-axis of the monoclinic setting below it. 
At cooling the sample from the paraelectric phase in presence of a low external field with $\varphi_E$ between $\varphi_1$ and $\varphi_3$ or between $\varphi_5$ and $\varphi_7$, which component along the $\mathbf{P}_{(1)}$--$\mathbf{P}_{(3)}$ axis  is larger than the tangential one, the (pseudo)spontaneous sublattice polarizations in the AFE* phase will be directed along this axis, and 1 and 3 configurations will be realized. However, the field with $\varphi_E$ between $\varphi_7$ and $\varphi_1$ or $\varphi_3$ and $\varphi_5$, where the tangential component to the $\mathbf{P}_{(1)}$--$\mathbf{P}_{(3)}$ axis is larger than the longitudinal one, changes that preferential direction for the spontaneous polarization/deformation of the sample, so the configurations 2 and 4 become the ground state ones, and the monoclinic axis  $a$ goes along $c_t$. 
 
 It should be kept in mind that the present theory does not describe the experimental situation, when the electric field is applied to the crystal, already spontaneously polarized  (antiferroelectrically) in the zero field, and the spontaneous proton configuration does not coincide with the one favored by the applied field. In this situation, the system finds itself in a metastable local minimum of the system energy, and with increasing the field, a transition to the global minimum will occur. Instead, we always consider the behavior of the system, which is in the global minimum of its thermodynamic potential. Then, the (pseudo)spontaneous proton configurations and the directions of the (pseudo)spontaneous sublattice polarizations are always those, which are dictated by the external electric field. The experimental procedure, which would reproduce this behavior, is to apply a small by magnitude field of the required direction to a heated crystal in the paraelectric phase, then cool the sample down to the required temperature, and then increase the field up to the required magnitude. 
 
 The lines II and III in figure~\ref{pdphi} are very well approximated by the piecewise functions  
 $E^{II}=E_{sw}/\cos(\Delta\varphi)$  and $E^{III}=E_{sw}/\sin(\Delta \varphi)$, where
 $\Delta \varphi=\varphi_E-\varphi_{2k+1}$, 
 $E_{sw}=214$~kV/cm,  and each piece is defined between  $\varphi_{2k}$ and $\varphi_{2k+2}$.
 At $\Delta \varphi=\pm\piup/4$, or, equivalently  $\varphi_E=\varphi_{2k}$, the fields $E^{II}$ and $E^{III}$
coincide, in accordance with figure~\ref{pdphi}.
A semi-qualitative explanation of these dependences is as follows. At zero temperature in the NC90 phase, the polarizations are oriented at $\piup/4-\Delta\varphi$ and $\piup/4+\Delta\varphi$ to the field  (see figure~\ref{circle}). In the FE phase, the sublattice polarizations are parallel, both at $\piup/4-\Delta\varphi$ to the field at zero temperature. In the AFE phase, the sublattice polarizations are antiparallel, directed at $\piup/4-\Delta\varphi$ and at $\piup-(\piup/4-\Delta\varphi)$  to the field. The difference between the energies of the fully ordered NC90 and FE arrangements, where two sublattices are treated as two dipoles in an external field, is proportional to  $(E \sin \Delta\varphi)$. On the other hand, the difference between the AFE and NC90 arrangements is proportional to  $(E \cos \Delta\varphi)$. When, roughly speaking, one of this differences becomes larger than the difference between the energies of the AFE correlations in the same arrangements, which is field-independent, the phase transition occurs. From here  directly follow the  above given dependences of the transition fields $E^{II}$ and $E^{III}$ on $\Delta\varphi$.

The  above discussion is relevant for low temperatures and for a fully ordered system, when all protons within each sublattice are arranged in the same lateral configuration. As a result, the sublattice polarization directions coincide with the directions of the total dipole moments of these configurations, and only four directions, along the $\mathbf{P}_{(1)}$--$\mathbf{P}_{(3)}$ and $\mathbf{P}_{(2)}$--$\mathbf{P}_{(4)}$ axes, are allowed. Thermal fluctuations in the presence of the electric field destroy this order, and some protons rearrange into other configurations, including the single-ionized ones, which dipole moments are directed at $45^\circ$ to these axes. In this case, the resulting \textit{average} vector of the sublattice polarization can be directed arbitrarily, as determined by relative populations of different proton configurations within the sublattice. Herein below we discuss the temperature--electric field behavior of squaric acid for some particular orientations of the field.

\subsection{Electric fields $\pm E_1$ and $\pm E_3$}
\begin{figure}[!t]
 	\centerline{\includegraphics[width=0.55\columnwidth]{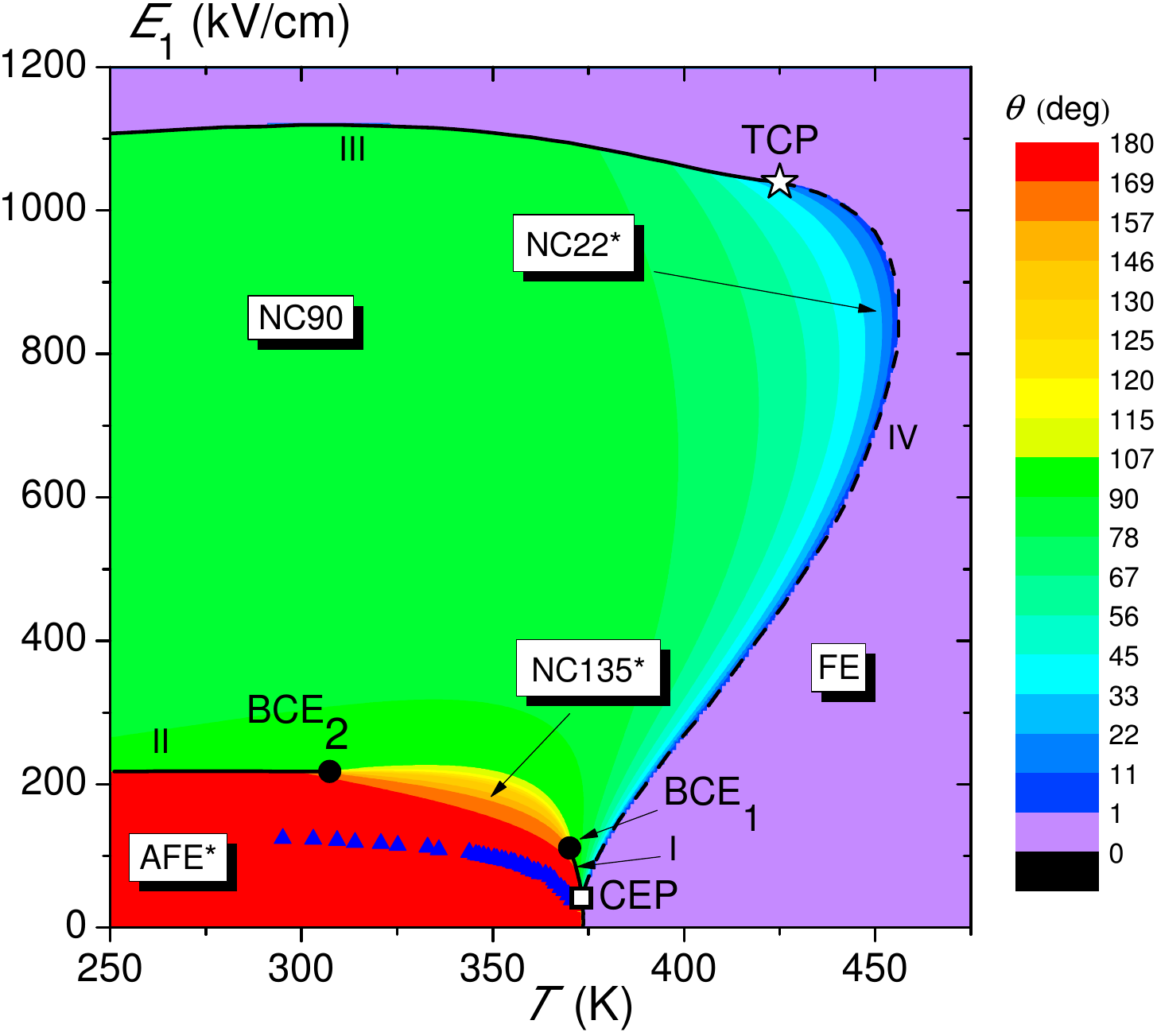}}
 	\caption{(Colour online) The $T$--$E_1$ phase diagram of the squaric acid, overlapping the $T$--$E_1$ color contour plot of the noncollinearity angle $\theta$. Solid and dashed lines indicate the first and second order phase transitions, respectively.  The open square {\large $\square$}, star \FiveStarOpen, and full circles {\large $\bullet$} indicate the critical end point, tricritical point, and bicritical end points, respectively. Blue full triangles $\blacktriangle$ are the experimental points \cite{horiuchi:18}.} \label{pdthetaE1}
 \end{figure}

These fields correspond to $\varphi_E=\piup n/2$.
As mentioned above, due to the symmetry of the system with respect to $90^\circ$  rotations around the $b$ axis, the $T$--$E$ phase diagrams for these fields will be identical, though the polarization vectors in all phases will be rotated counterclockwise by $\piup n/2$ as compared to the case of the field $E_1$. The $T$--$E_1$ diagram of squaric acid was obtained earlier \cite{moina:21}. This diagram, overlapped with the color gradient plot of the noncollinearity angle $\theta$, is shown in figure~\ref{pdthetaE1}.  We detected three lines of the first order phase transitions I, II, and III. Lines I and II terminate at the bicritical end points BCE$_1$ and BCE$_2$. Line III ends at the tricritical point TCP. The line of the second order phase transitions IV starts at the tricritical pont TCP and terminates at the critical end point CEP, lying at the line of the first order transitions I in a close vicinity of the bicritical end point BCE$_1$.

The phase denoted as AFE* (the red region) is non-collinear antiferrielectric, but very close to the initial AFE phase. Here, $\theta\sim 180^\circ$, as only a slight rotation of the minus sublattice polarization takes place in this region at non-zero fields, discernible mostly at its upper right boundary.
The purple region on the phase diagram is the collinear field-induced ferroelectric phase (FE) with $\theta=0$. The true paraelectric phase with zero sublattice polarizations, just like the true AFE phase, exists only at $E=0$.  
The phase between the transition lines II, III, and IV (the green and blue regions in figure~\ref{pdthetaE1}) is the non-collinear ferrielectric phase NC90. In this phase, the noncollinearity angle $\theta$ is mostly close to 90$^\circ$, only rapidly decreasing to zero at increasing temperature within a narrow (blue) stripe along the second-order phase transition line IV, denoted as NC22*.

In the region NC135* between the BCE$_1$ and BCE$_2$ points (orange to yellow in figure~\ref{pdthetaE1}) a crossover between the AFE* and NC90 phases occurs. Here,  $\theta$  gradually changes from $\sim180^\circ$ to $\sim90^\circ$: the minus sublattice polarization rotates continuously with increasing field and becomes perpendicular to the plus sublattice polarization. As discussed above, this continuous rotation in the NC135* and NC22* regions is a statistically averaged effect, possible only in the presence of thermal fluctuations. 

As one can see in figure~\ref{pdthetaE1}, the calculated fields of switching between the AFE* and NC90 phases, which occurs either as a phase transition at lines II and I or as a crossover within the region NC135* between the points BCE$_1$ and BCE$_2$, are essentially higher than the ones observed experimentally \cite{horiuchi:18}.

\subsection{Diagonal electric fields $E_1\pm E_3$}
These are the fields directed along the diagonals of the tetragonal $ac$ plane: $\varphi_E=\piup/4+\piup n/2$. For the sake of convenience, we  denote them as $E_1\pm E_3$. For all these fields, the $T$--$E$ phase diagrams are identical because of the
the symmetry of the system with respect to $90^\circ$  rotations around the $b$ axis.

	\begin{figure}[!t]
\centerline{
 \includegraphics[height=0.55\textwidth]{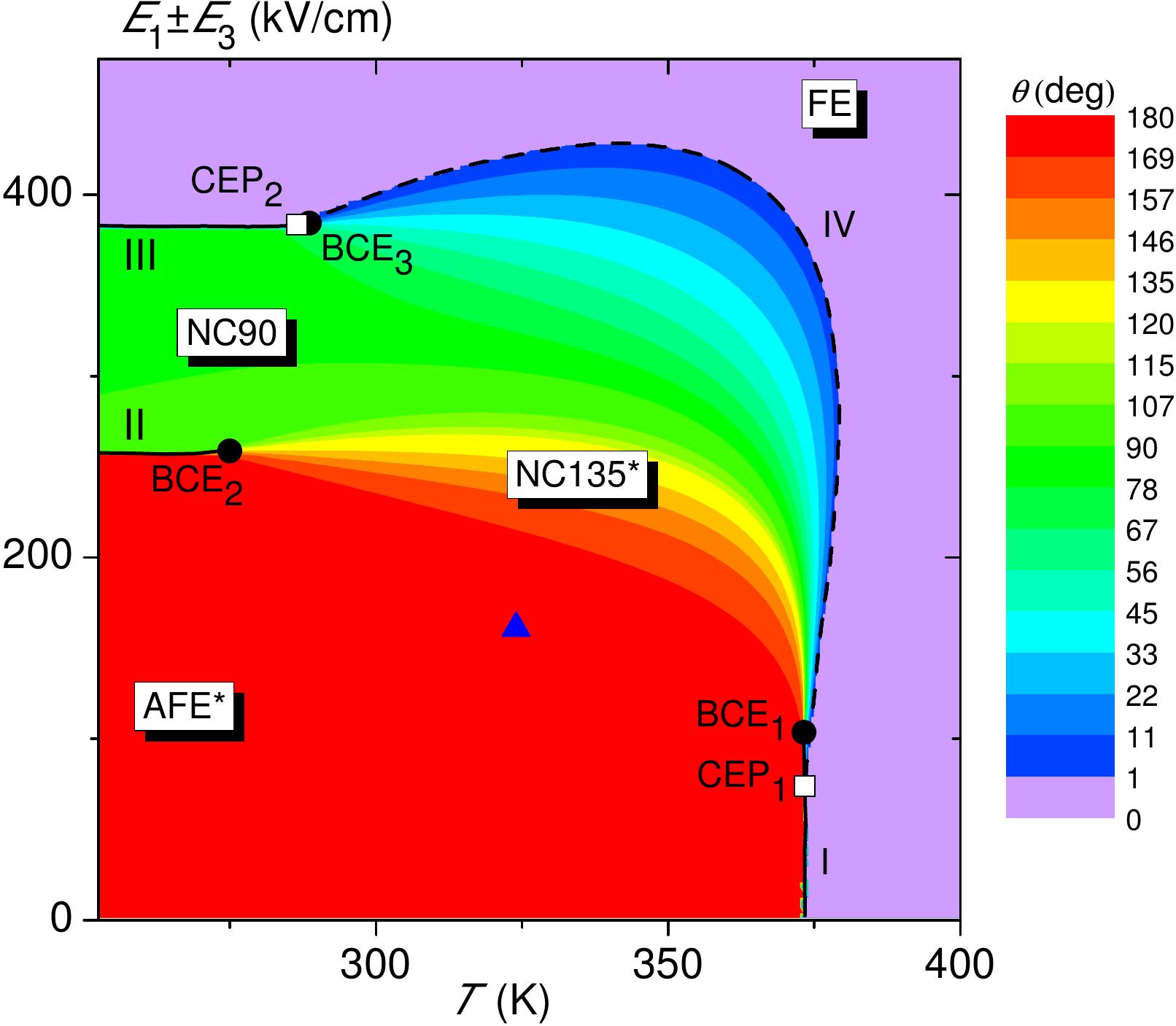}  		\hspace{3ex}	\includegraphics[height=0.58\textwidth]{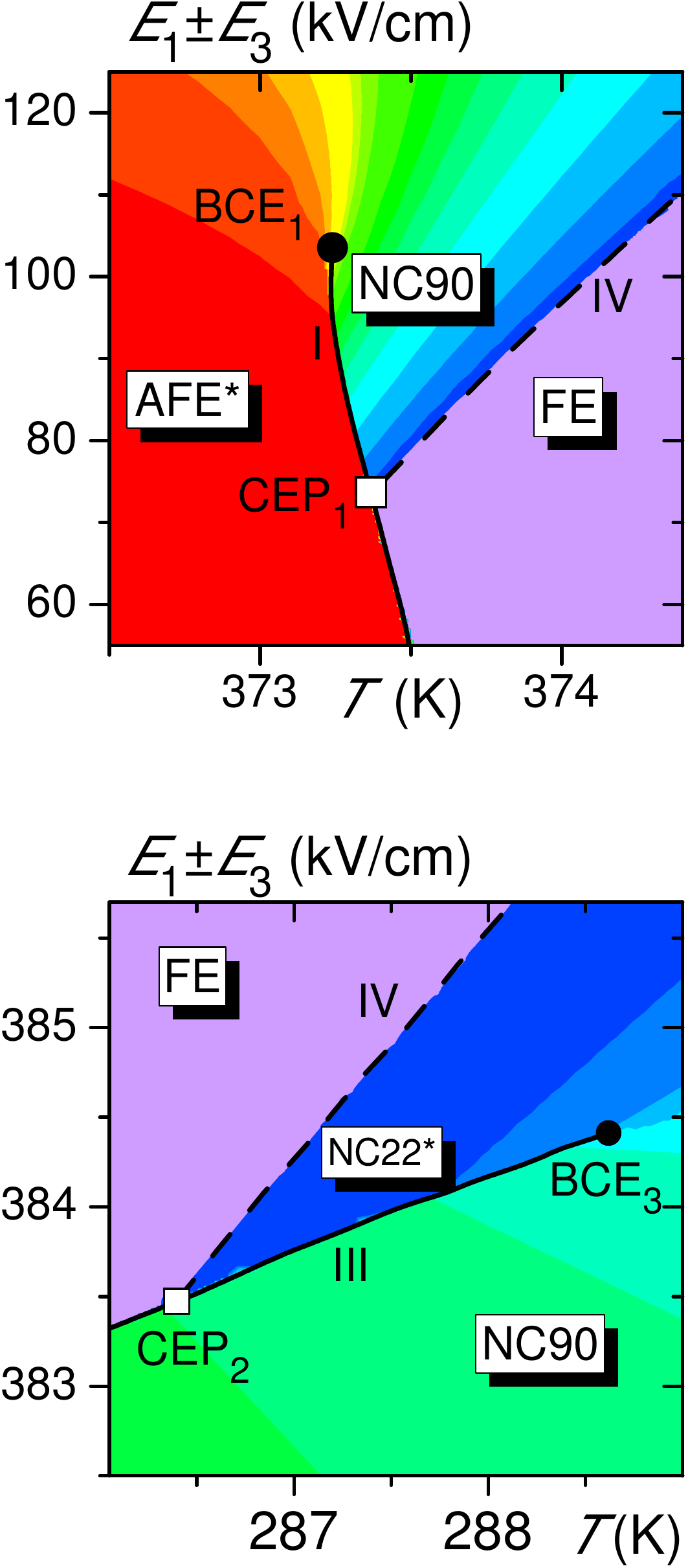}
}
\caption{(Colour online) The same as in figure~\ref{pdthetaE1} for the fields $E_1\pm E_3$. The tricritical point TCP is split to the critical end point CEP$_2$ and bicritical end point BCE$_3$.  Graphs in the right-hand column are the closeups of the phase diagram in the vicinities of BCE$_1$/CEP$_1$ and BCE$_3$/CEP$_2$.  Blue full triangle $\blacktriangle$ is the experimental point obtained for $E_1+E_3$ (taken from the Electronic supplementary material to \cite{horiuchi:18}).} \label{pdthetaE1m3}
\end{figure}

As one can see in figure~\ref{pdthetaE1m3}, the $T$--$(E_1\pm E_3)$ phase diagram topology is mainly similar to that of the $T$--$E_1$.  The fields of the low-temperature switching to the noncollinear phase NC90 (line II) are higher than predicted for the field $E_1$, but the fields of the transition to the collinear FE phase (line III) are much lower. 
The increase of the switching field $E^{II}$ for  $E_1\pm E_3$ as compared to the case of $E_1$ is in a qualitative agreement with the experimental observations (see \cite{horiuchi:18} and the Electronic supplementary material to it). The theoretical values of $E^{II}$ are, however, significantly higher than the measured value.

There are also some qualitative differences between
the phase diagrams, predicted for the fields $E_1$ and $E_1\pm E_3$.
The obvious difference  is that the tricritical point TCP from the $T$--$E_1$ diagram is here split into the system of the bicritical end point BCE$_3$, terminating the line III of the first order transitions, and the critical end point CEP$_2$, where line IV of the second order transitions starts (see figure~\ref{pdthetaE1m3}).  
In the emerged wedge between the lines III and IV, a new phase
NC22* is stabilized. This is a noncollinear ferrielectric phase
with a small angle $\theta$ between the sublattice polarizations (see figure~\ref{pdthetaE1m3}). 
It becomes clear then that in the $T$--$E_1$ diagram (figure~\ref{pdthetaE1}), the region NC22* is a prototype of the NC22* phase, not fully stabilized at the chosen values of the model parameters.

Such a splitting of a TCP into a system of a BCE and CEP points is a well known phenomenon in the theory of metamagnetic transitions, occurring, for instance, in the two-sublattice Ising model for some sets of the model parameters, in particular, at different ratios between the competing intra- and inter-sublattice interactions \cite{kincaid:75}. Even though we use the same set of the model parameters for all $\varphi_E$, changing the field from $E_1$ to $E_1\pm E_3$  is, technically, equivalent to changing the values of the dipole moments $\mu^H+\mu^\pi_\parallel$ and $\mu^\pi_\perp$, as can be seen from equations (\ref{z}) and (\ref{dipole1}), (\ref{dipole3}). No attempt has been made to determine the direction of the field, at which this splitting of the TCP point appears in the phase diagram. Moreover, the splitting can appear even in the $T$--$E_1$ diagram, if other values of the parameters are used. This is another ramification of the uncertainty in the choice of the model parameters.

 There are indications that for the two-sublattice Ising and similar models, the splitting of the TCP point is an artefact of the used mean field approximation (see e.g., \cite{plascak:03} and references therein) and is never realized in experiment. Nonetheless, it clarifies some peculiarities of the system behavior in a situation, when the splitting does not occur with the present set of the model parameters, like the discussed earlier case of the field $E_1$. 
\newpage

\subsection{Fields directed at $\piup/4$  to the sublattice polarization axes}
Those  are the fields directed at  $\varphi_E=\varphi_{2k+1}$ to the $a$ axis (e.g. $\varphi_7\approx 11^\circ$ for the used set of the model parameters). Since at these $\varphi_E$,  the switching of AFE proton configurations  extensively discussed in 
subsection~\ref{groundstate} takes place, in order to avoid confusion, we shall imply that
the case of $\varphi_E=\varphi_{2k+1}-\Delta \varphi$,
$\Delta\varphi\to  0$, is considered.


The phase diagram at low fields is very similar to the $T$--$E_1$ one. The AFE, NC90, and FE phases are observed. The BCE$_1$ and BCE$_2$ points are closer here than in the field $E_1$. No second, high-field transition from NC90 to the NC22* or FE phase at low temperatures is detected below 10$^9$~kV/cm.

\begin{figure}[!t]
	\centerline{	\includegraphics[height=0.4\textwidth]{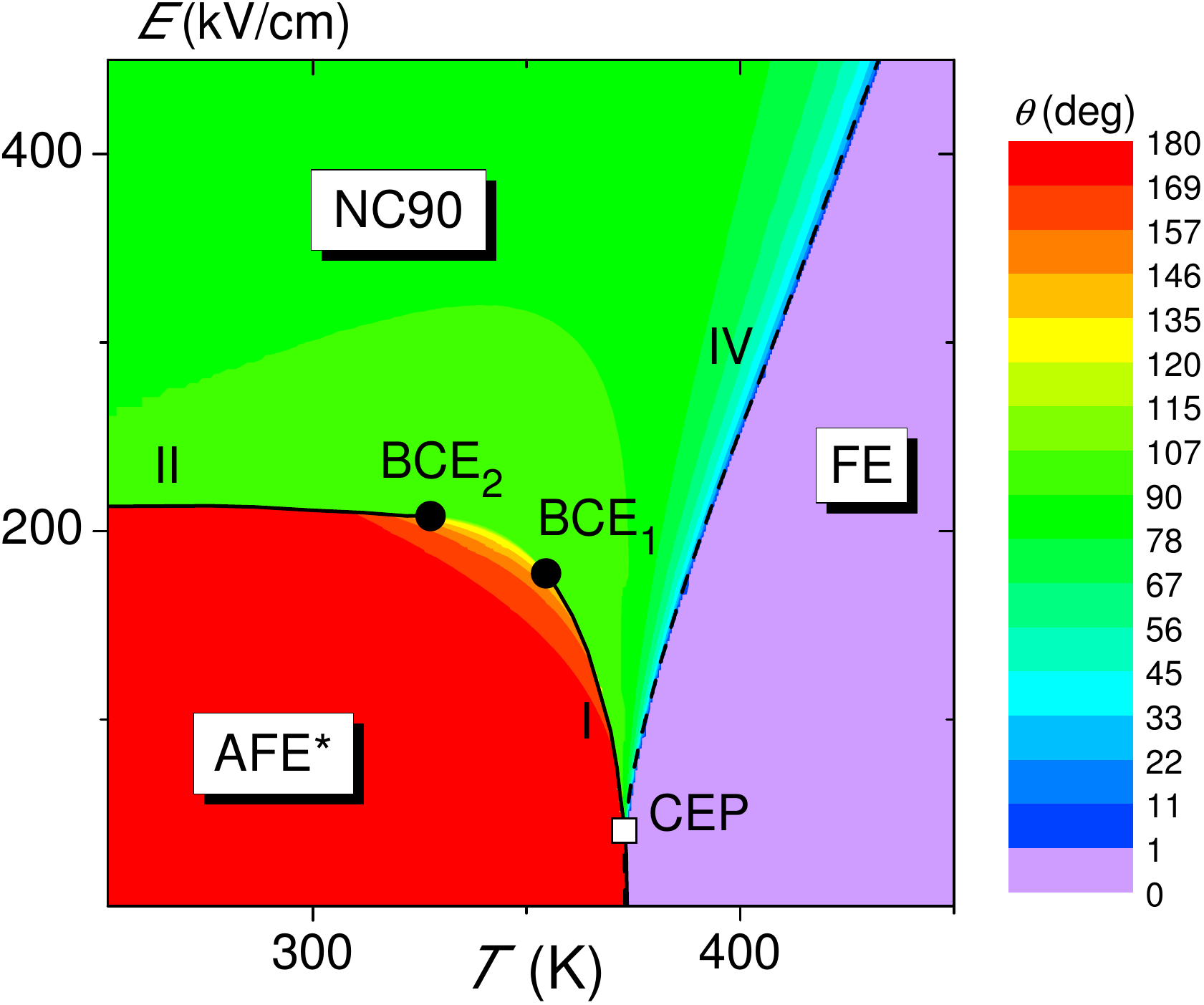} }
	\caption{(Colour online) The same as in figure~\ref{pdthetaE1} for the fields directed at $\varphi_E=\varphi_{2k+1}$ to the $a$ axis. } \label{pd11}
\end{figure}

\subsection{Fields collinear to the sublattice polarizations }
\begin{figure}[!t]
		\centerline{	\includegraphics[height=0.45\textwidth]{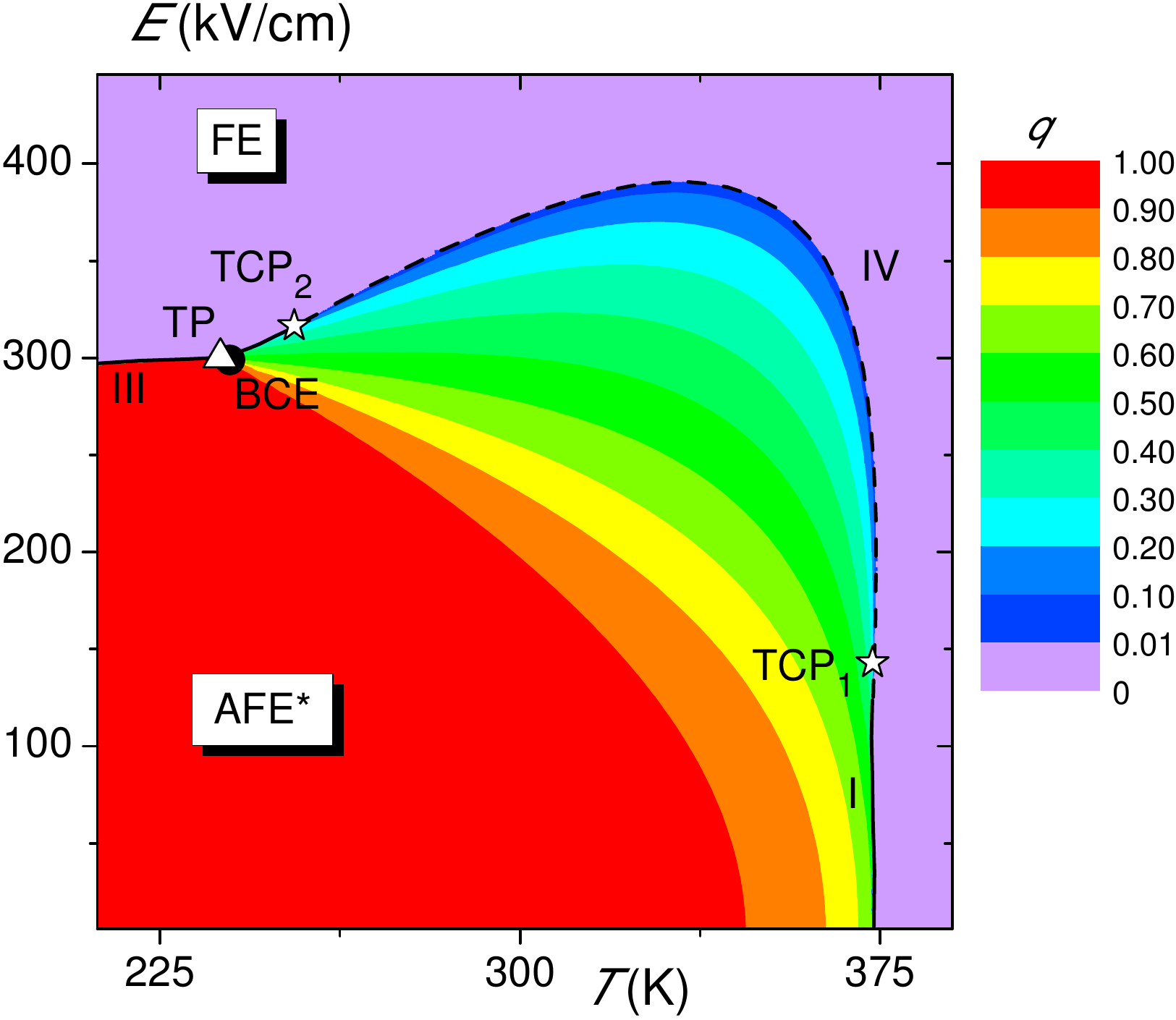} \hspace{2ex}
		\includegraphics[height=0.35\textwidth]{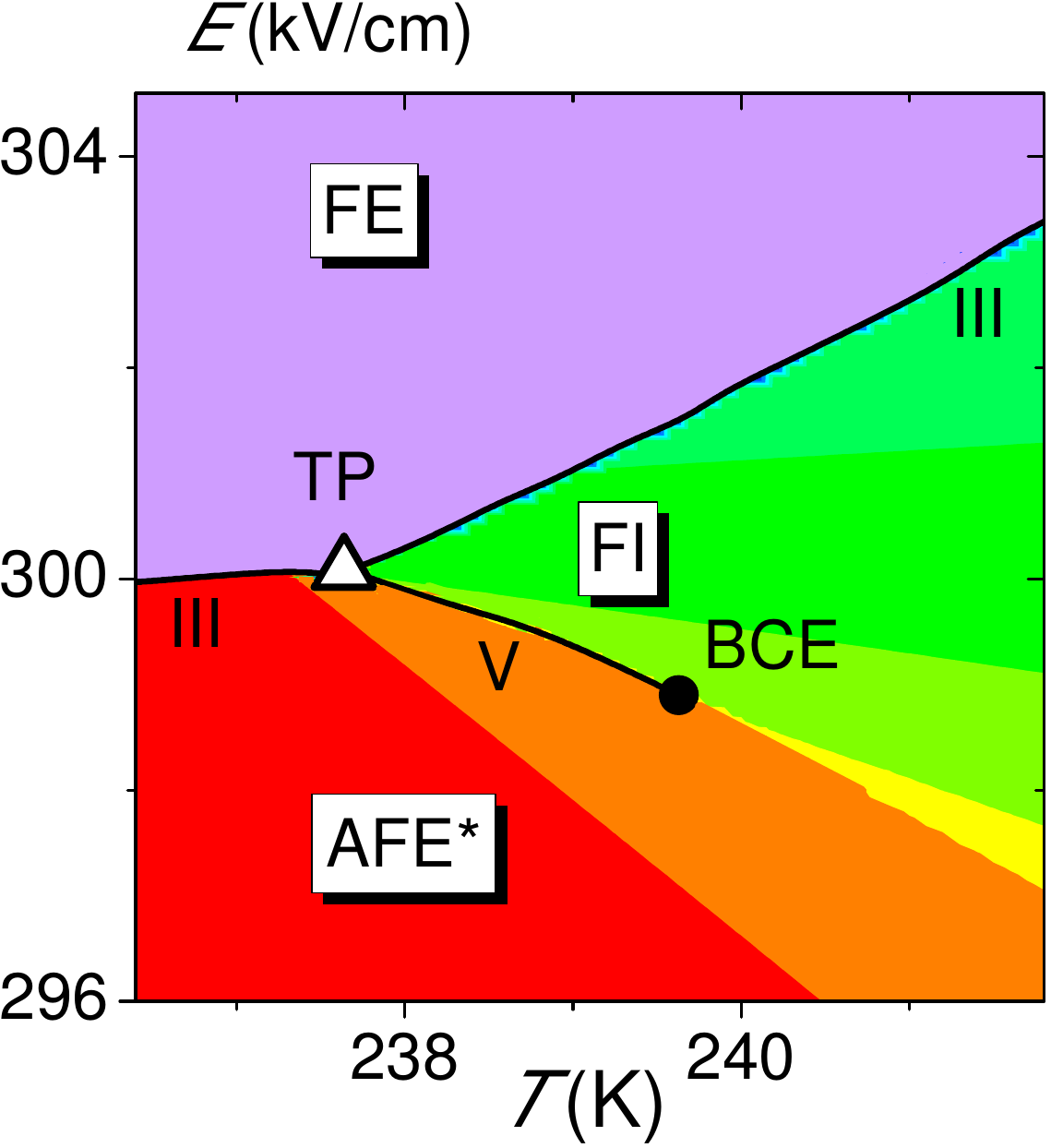}}
	\caption{(Colour online) The $T$--$E$ phase diagram of the squaric acid, overlapping the $T$--$E$ color contour plot of the AFE ordering parameter $q$ for the field collinear to the polarization axis ($\varphi_E=\varphi_{2k}$). The right-hand panel: a closeup of the phase diagram in the vicinity of the triple point TP. Solid and dashed lines indicate the first and second order phase transitions, respectively.  The  stars \FiveStarOpen, full circles {\large $\bullet$}, and open triangles $\triangle$ indicate the tricritical points, bicritical end point, and the triple point, respectively. } \label{pd56}
\end{figure}

Those are the fields directed at $\varphi_E=\varphi_{2k}$ (e.g. $\varphi_0\approx 56^\circ$). Being collinear to the sublattice polarization vectors in the AFE phase, they have no tangential component to rotate these vectors within the $ac$ plane. The squaric acid crystal then behaves like a uniaxial antiferroelectric, where the field-induced switching occurs by flipping the minus sublattice polarization along the same direction. The switching at low temperatures is thus  a one-step process from the AFE* to the FE phase. At all temperatures and field magnitudes $\eta_{1\pm}=-\eta_{2\pm}$. In this case, the noncollinearity angle $\theta$ cannot be used to distinguish between different phases, as it always equals either $180^\circ$  or 0. Instead, we use the traditional AFE ordering parameter
$q=\eta_{2+}-\eta_{2-}$. The obtained $T$--$E$ phase diagram, overlapping the color contour plot of $q$ is shown in figure~\ref{pd56}.

The line II and BCE$_2$ point are obviously absent from the diagram. At higher temperatures, the 
intermediate collinear ferrielectric phase FI is observed between the AFE* and FE phases.  The FI phase is, apparently, an analog of the NC22* phase/region, observed 
in the phase diagrams for the fields, noncollinear to the polarization axis.  In the AFE* and FI phases $\eta_{f+}=-\eta_{f-}$, in the FE phase $\eta_{f+}=\eta_{f-}$. 

The system of the bicritical end point BCE$_1$ and critical endpoint CEP$_1$, observed at the $T$--$E_1$ and $T$--$(E_1\pm E_3)$ phase diagrams, collapses here into a single tricritical point TCP$_1$, at which the line I of the first order transitions ends, and the line IV of the second order transitions starts. Line IV continues up to the other tricritical point TCP$_2$, at which the line III of the first order transitions begins. Not far from the TCP$_2$ point there is a triple point TP, at which the AFE*, FE, and FI phases are in equilibrium. The line V of the first order transitions, separating the AFE* and FI phases, is very short, starting from the triple point TP and ending at the bicritical end point BCE. A downward jump of $q$ occurs, when temperature is increased across line V. The phase diagram, overall, is highly reminiscent of the one obtained by the Landau expansion method for uniaxial antiferroelectrics \cite{suzuki:83} (figure~1b therein), especially in the region of the TP-TCP$_2$-BCE points.

\section{Concluding remarks}
Using the previously developed \cite{moina:21} deformable two-sublattice proton ordering model, we explore the processes of polarization rotation in antiferroelectric crystals of squaric acid under the inluence of external electric fields, oriented arbitrarily within the plane of the hydrogen bonds. The unique structure of the two-dimensional hydrogen bond networks in squaric acid permits 90$^\circ$ rotation of the sublattice polarization. Application of the model to the case of the electric field $E_1$  has revealed \cite{moina:21} that unlike in the conventional uniaxial antiferroelectrics, the polarization rotation in squaric acid is a two-step process, as it was predicted earlier \cite{horiuchi:18}. At low temperatures the crystal is switched from the antiferroelectric  phase AFE*, first to the noncollinear phase NC90 with perpendicular sublattice polarizations and only then to the collinear ferroelectric phase FE.

From the low-temperature $\varphi_E$--$E$ phase diagram constructed in the present paper  where $\varphi_E$ is the angle between the field and the $a$ axis, it follows that the polarization rotation is the same two-step process for all $\varphi_E$ except for a few particular directions with $\varphi_E=\varphi_n$, where $\varphi_n=\varphi_0+\piup n /4$, $n=0,...7$. For these special directions, the field is either collinear to the axes of the sublattice polarization in the AFE* phase (even $n$, $\varphi_E=\varphi_{2k}$) or directed at $\piup/4$ to these axes,
(odd $n$, $\varphi_E=\varphi_{2k+1}$). For $\varphi_E=\varphi_{2k}$,
the crystal behaves like a uniaxial antiferroelectric, undergoing a single-step polarization switching to the FE phase without the intermediate noncollinear phase.
On the other hand, for $\varphi_E=\varphi_{2k+1}$, the transition field from the NC90 to the FE phase tends to infinity, meaning that the transition never occurs. The dependences of the first-order transition fields $E^{II}$ and $E^{III}$ between the AFE* and NC90 and between the NC90 and FE phases, respectively, on $\varphi_E$ are given by the piecewise functions 
$E^{II}=E_{sw}/\cos(\Delta\varphi)$  and $E^{III}=E_{sw}/\sin(\Delta \varphi)$, where
$\Delta \varphi=\varphi_E-\varphi_{2k+1}$, 
 and each piece is defined between  $\varphi_{2k}$ and $\varphi_{2k+2}$.

Experimentally, the low-temperature transition to the ferroelectric phase with increasing field has not been detected yet because of the dielectric breakdown of the samples. As follows from the constructed $\varphi_E$--$E$ diagram, the field of this transition is the lowest, when its direction is close to the axis of the sublattice polarization, $\varphi_E=\varphi_{2k}$, so it is most likely to be experimentally observed at this field orientation.  
%

The $T$--$E$ phase diagrams are constructed for the electric fields, directed along the diagonals of the $ac$ plane, for the fields collinear to the axes of the sublattice polarization in the AFE* phase, $\varphi_E=\varphi_{2k}$, as well for the fields directed at $\piup/4$ to these axes $\varphi_E=\varphi_{2k+1}$.

\newpage
\section*{Appendix}
	\begin{figure}[!h]
	\begin{center}
		
\begin{flushleft}{$\varphi_5<\varphi_E<\varphi_7$}\end{flushleft}
		\centerline{\includegraphics[width=0.80\textwidth]{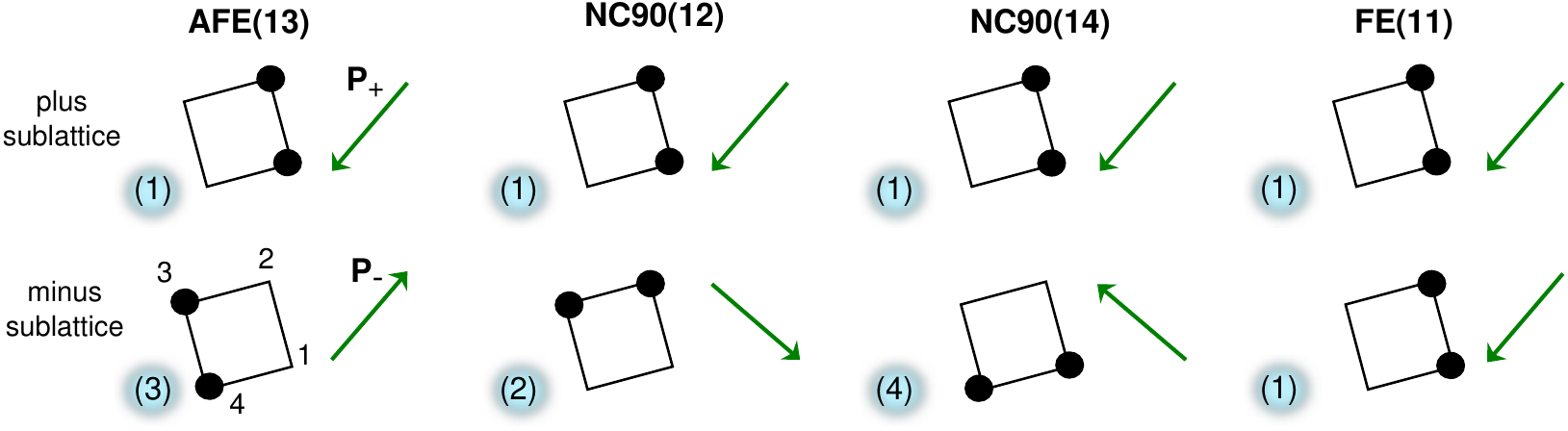}} 
		\rule{0.85\textwidth}{0.4pt}
		\smallskip
		
\begin{flushleft}$\varphi_7<\varphi_E<\varphi_1$\end{flushleft}
		\centerline{\includegraphics[width=0.80\textwidth]{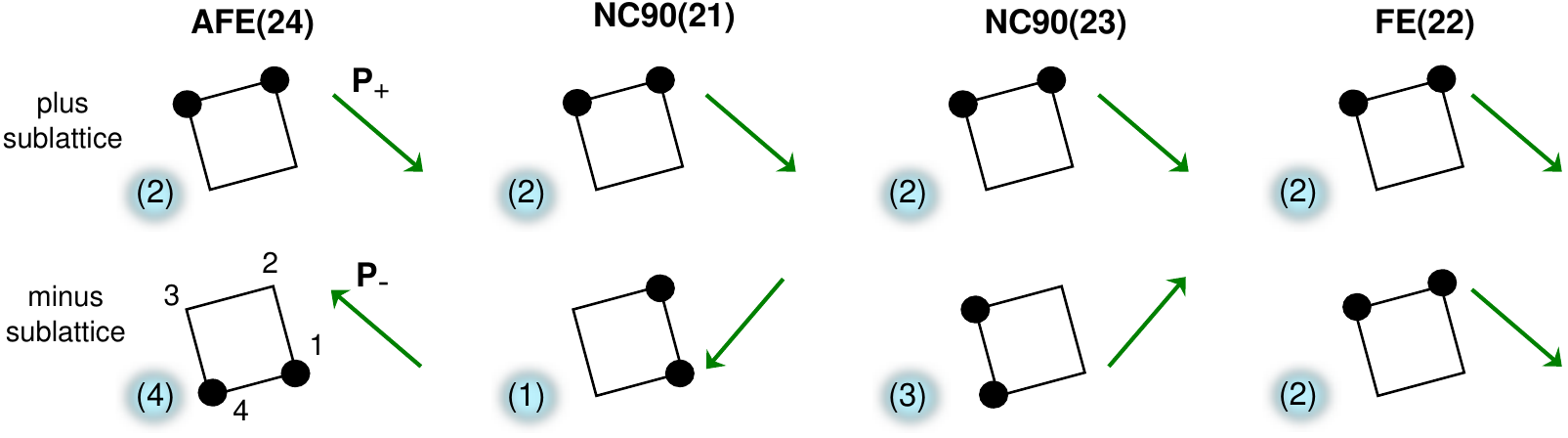}}
		\rule{0.85\textwidth}{0.4pt}

		\smallskip
		
\begin{flushleft}$\varphi_1<\varphi_E<\varphi_3$\end{flushleft}
		\centerline{\includegraphics[width=0.80\textwidth]{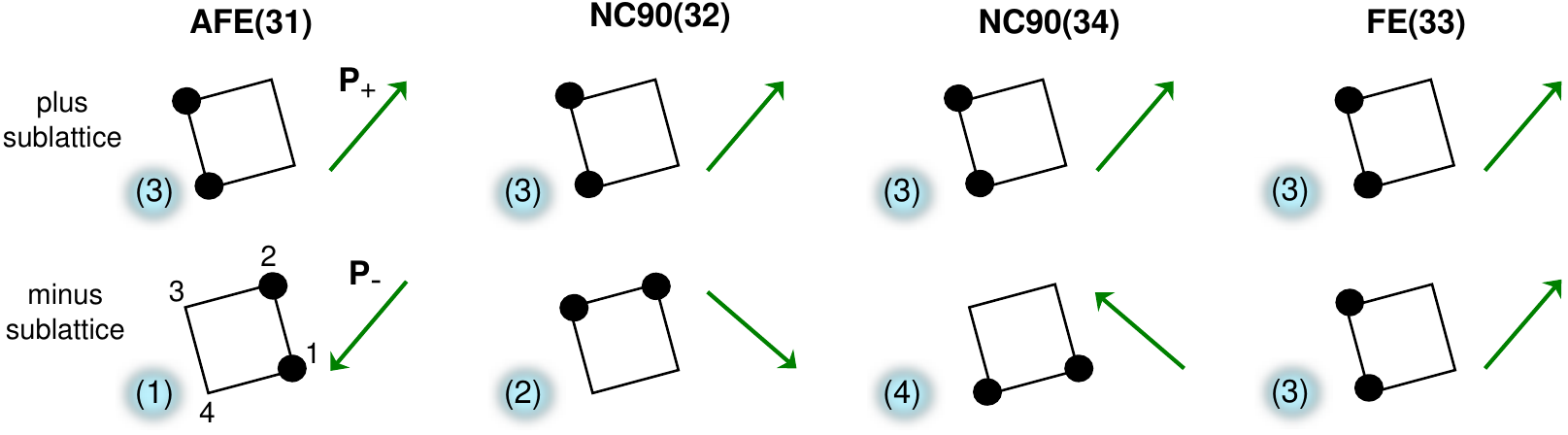}}		
		\rule{0.85\textwidth}{0.4pt}
		
		\smallskip
		
\begin{flushleft}$\varphi_3<\varphi_E<\varphi_5$\end{flushleft}
		\centerline{\includegraphics[width=0.80\textwidth]{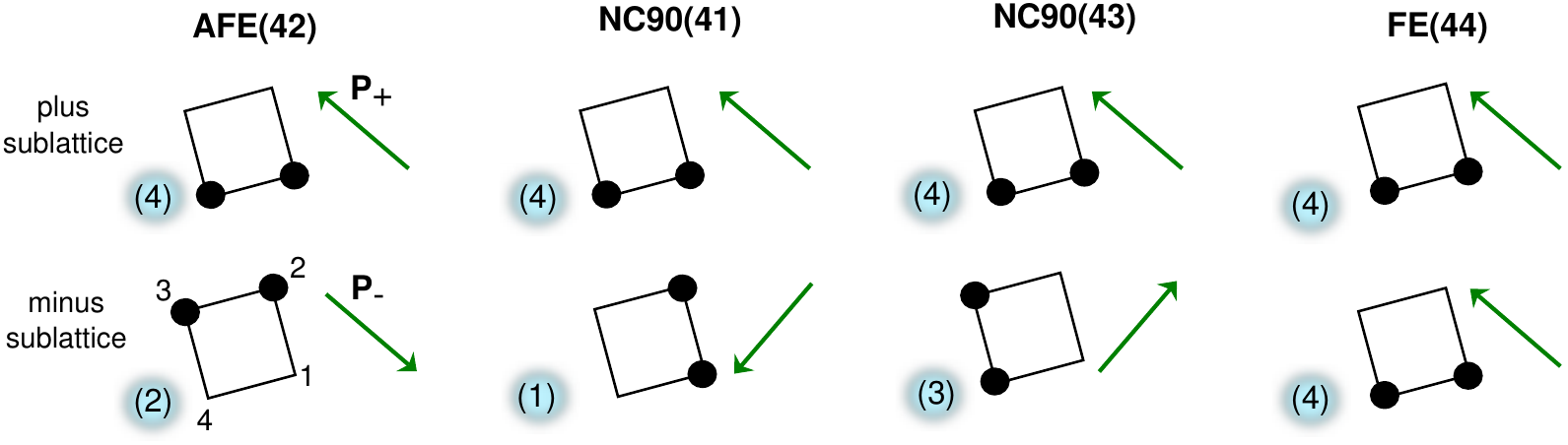}}
				\end{center}
	\caption{(Colour online) The ground state proton configurations realized in different phases at different orientations of the electric field (see figure~\ref{circle}). The polarization vectors $\mathbf{P}_+$ and $\mathbf{P}_{-}$ of the plus and minus sublattices are shown by the green arrows. The numbers in parentheses are the numbers of the respective lateral proton configurations from table~\ref{configurations_table}. The hydrogen bond indices $f=1,2,3,4$ are shown.  }
	\label{phasesE1}
\end{figure}

\section*{Note added in proof}
The recently published results~\cite{horiuchi:21} of
of the revisited hysteresis loops measurements in the squaric
acid crystals with improved dielectric strength appear to be much better described by the present
theory than the earlier experimental data of \cite{horiuchi:18}. 
%
%
The difference between the theoretical and experimental values of the
switching field $E^{II}_1$ at room temperature reduces twice.
The agreement with the experiment for
the polarization values in the NC90 phase induced by the field $E_1$ is significantly improved as
well. It renders a simultaneous quantitative description of experimental data for the permittivity,
polarization, and switching fields by the present model with the same set of the fitting parameters
though not attained yet, but much more feasible than it was previously thought.

\newpage
\ukrainianpart

\title{Вплив напрямку зовнішнього поля на обертання поляризації   в антисегнетоелектричних кристалах квадратної кислоти H$_2$C$_4$O$_4$}
\author{А. П. Моїна}
\address{
	Інститут фізики конденсованих систем Національної академії наук України, \\вул. Свєнціцького, 1, 79011 Львів, Україна
}
%
%
%

\makeukrtitle

\begin{abstract}
	\tolerance=3000%
	З використанням запропонованої раніше моделі досліджуються процеси обертання поляризації зовнішніми електричними полями, орієнтованими довільним чином в площині $ac$, в антисегнетоелектричних кристалах квадратної кислоти.
	Передбачається, що за винятком декількох особливих напрямків поля, реорієнтація поляризації при низьких температурах відбувається в два етапи: спершу до неколінеарної фази з перпендикулярними поляризаціями підграток, а потім до колінеарної сегнетоелектричної фази. Однак, коли поле орієнтоване уздовж осі спонтанних поляризацій підграток, проміжна неколінеарна фаза відсутня. Коли ж поле спрямоване під кутом 45$^\circ$ до цієї осі, то величина поля, при якому відбувається перехід до сегнетоелектричної фази, прямує до нескінченності. Для усіх напрямків зовнішнього поля визначені протонні конфігурації основного стану та напрямки векторів поляризації підграток. Побудовано $T$--$E$ фазові діаграми для полів, спрямованих уздовж діагоналей площини $ac$, а також для полів згаданих особливих напрямків.
	
	\keywords
	антисегнетоелектрик, електричне поле, фазовий перехід, фазова діаграма, квадратна кислота
	
\end{abstract}

\end{document}